\documentclass[aps,prl,reprint,superscriptaddress]{revtex4-2}
\usepackage{graphicx} 
\usepackage{float}
\usepackage{amsmath} 
\usepackage{hyperref}
\usepackage{physics}
\usepackage{mathtools}
\hypersetup{
        colorlinks,
   citecolor=blue,
   filecolor=blue,
   linkcolor=blue,
   urlcolor=blue,
   breaklinks=true}
\usepackage{xcolor}

\newcommand{\kB}{k_\text{B}}

\begin{document}
\title{Origin of Edge Currents in Chiral Active Liquids}
\author{Faisal Alsallom}
\email{alsallom@berkeley.edu}
\affiliation{Department of Physics, University of California, Berkeley, California 94720, USA}
\affiliation{Chemical Science Division, Lawrence Berkeley National Laboratory, Berkeley, California 94720, USA}

\author{David T. Limmer}
\email{dlimmer@berkeley.edu}
\affiliation{Department of Chemistry, University of California, Berkeley, California 94720, USA}
\affiliation{Materials Science Division, Lawrence Berkeley National Laboratory, Berkeley, California 94720, USA}
\affiliation{Chemical Science Division, Lawrence Berkeley National Laboratory, Berkeley, California 94720, USA}
\affiliation{Kavli Energy NanoSciences Institute, Berkeley, California 94720, USA}

\date{\today}

\begin{abstract}
Chiral active liquids generically exhibit unidirectional edge currents in confinement. While this phenomenon has been attributed to model-specific mechanisms and interpreted through phenomenological equations, a universal understanding of it, and its connection to microscopic dynamics remain absent. Starting from the microscopic equations of motion of a simple interacting two-dimensional model, we find that localized edge currents emerge as a consequence of global angular momentum balance. From these underlying equations, we derive an Ohmic-like conductance law for the mean edge current in the dense phase, and we find it to be intensive, depending only on the density, active torque and substrate drag. For simple geometries, we find the distribution of the edge currents has a closed Gaussian form, with a variance that is intensive, depending only on temperature, density and the aspect ratio of the system. These results are validated numerically using extensive molecular dynamics simulations. This origin of the edge current is shown to extend to other models of chiral systems where angular momentum is injected in distinct ways.

\end{abstract}

\maketitle

Chiral active systems are composed of constituents that actively generate torques and rotations by consuming energy from their environments, driving them out of equilibrium and breaking time-reversal and parity symmetries~\cite{bowick2022symmetry}. These systems host  exotic transport phenomena in their bulk, such as odd viscosity \cite{epstein2020time,Vitelli2017,Vitelli2021}, odd diffusivity \cite{Hargus2021}, odd mobility \cite{Poggioli2023_Mobility,hargus2025odd} and odd elasticity \cite{scheibner2020odd,Fakhri2022}. In addition to the local odd behavior of these systems, they also exhibit remarkable collective phenomena in the vicinity of their boundaries, such as spontaneously forming unidirectional persistent edge currents and Kelvin waves~\cite{Poggioli2023_Kelvin,soni2019odd,Langford2025,Benjamin2016,Siebers2024,Caporusso2024,tsai2005chiral,li2024robust,yang2020robust,yashunsky2022chiral,liu2020oscillating,metzger2026equation}, akin to the topologically protected edge modes in the quantum Hall effect \cite{girvin2019modern}. These boundary currents are forbidden in equilibrium, yet they are allowed in active systems as a result of broken detailed balance. Understanding this nonequilibrium collective behavior is challenging since the traditional formalism for coarse-graining particle dynamics into effective fields predominantly requires assumptions of local equilibrium that are violated in active fluids~\cite{marchetti2013hydrodynamics}. While previous studies have successfully devised hydrodynamic equations to describe the localized average velocity profile \cite{Poggioli2023_Kelvin,soni2019odd,Caporusso2024,tsai2005chiral,yang2020robust,li2024robust,yashunsky2022chiral,liu2020oscillating}, these approaches mostly depend on phenomenological constants, and are unable to connect these edge currents to microscopic parameters of the system.  However, a unifying physical origin of these edge currents beyond their system-specific mechanisms, as well as their connection to microscopic parameters remains unclear, with some studies suggesting their robustness to be topological  \cite{Dasbiswas2018,sone2020exceptional}. 

Here we demonstrate that the global balance of angular momentum provides an unambiguous origin for edge currents as well as a direct connection to the microscopic details of the system. We first derive effective hydrodynamic equations that explain the localization length of the current as a balance of dissipation through the substrate friction and internal shear viscosity. However, the total current emerges as a consequence of the accumulation of the injected angular momentum in the form of orbital angular momentum and requires a particle framework to describe. Then, starting from microscopic equations of motion for chiral active dimers, we derive a relationship between the global angular momentum, mean edge current, and microscopic system parameters, leading to a chiral edge current conductance law analogous to Ohm's law. This framework and analogous conductance laws are expected to be general for interparticle interactions that conserve angular momentum, as we analytically show in the supplementary materials (SM) for other chiral systems where angular momentum is introduced in distinct ways. Additionally, for the model we consider here, we can solve for the full statistics of the edge currents in simple geometries. These statistics are equilibrium-like, with activity shifting only the mean of the distribution.

We consider a minimal model of chiral active particles composed of $N$ identical dimers in two dimensions \cite{Hargus2020}. Every dimer consists of two equivalent monomers, each of mass $m$, bonded harmonically with a spring of stiffness $k$ and resting length $\sigma$. The bonding potential energy within a dimer is given by $k(\abs{\textbf{d}_i}-\sigma)^2/2$, where $\textbf{d}_i$ is the relative position vector connecting the two bonded monomers, and we will work in the limit that $k$ is large. The interaction potential energy between non-bonded monomers is given by the Weeks-Chandler-Andersen short-ranged repulsive interaction potential \cite{Weeks1971_WCA}, which sets their diameter to $\sigma$ and characteristic energy scale to be $\epsilon$. We evolve our system according to the under-damped Langevin equations of motion
\begin{equation}\label{Langevin}
    \begin{aligned}
\dot{\textbf{r}}_{i\alpha}&=\frac{\textbf{p}_{i\alpha}}{m},\\
        \dot{\textbf{p}}_{i\alpha}&=-\frac{\gamma}{m}\textbf{p}_{i\alpha}+\textbf{F}_{i\alpha}^c+\textbf{F}_{i\alpha}^a +\textbf{F}_{i\alpha}^\text{w}+\sqrt{2\gamma \kB T} \boldsymbol{\eta}_{i\alpha}(t),
    \end{aligned}
\end{equation}
where $i\in\{1,\dots,N\}$ indexes the dimer and $\alpha \in\{1,2\}$ indexes the monomer. The position and momentum vectors on the $i\alpha$th particle are denoted $\textbf{r}_{i\alpha}$ and $\textbf{p}_{i\alpha}$, respectively, $\textbf{F}_{i\alpha}^c$ is the total force due to potential forces acting on the $i\alpha$th particle, and $\textbf{F}_{i\alpha}^\text{w}$ is the force exerted on it by the wall. The active force acting on the $i\alpha$th particle is $\textbf{F}_{i\alpha}^a$. Within each dimer, each monomer experiences equal and opposite active forces, such that the forces are perpendicular to the instantaneous bonding vector $\textbf{d}_i$.  The active torque $\tau_a$ on each dimer is constant and equal, $F_{i\alpha}^a(t) d_i(t)=\tau_a$.  We consider these particles supported on a substrate with friction coefficient $\gamma$ and held at temperature $T$, with $\boldsymbol{\eta}_{i\alpha}$  a Gaussian delta-correlated random noise with zero mean and unit variance \cite{limmer2024statistical}. In confined geometries, we model walls as frictionless with a short-range repulsive potential. Details on the simulations can be found in the SM.

\begin{figure}[t]
    \centering
    \includegraphics[width=\columnwidth]{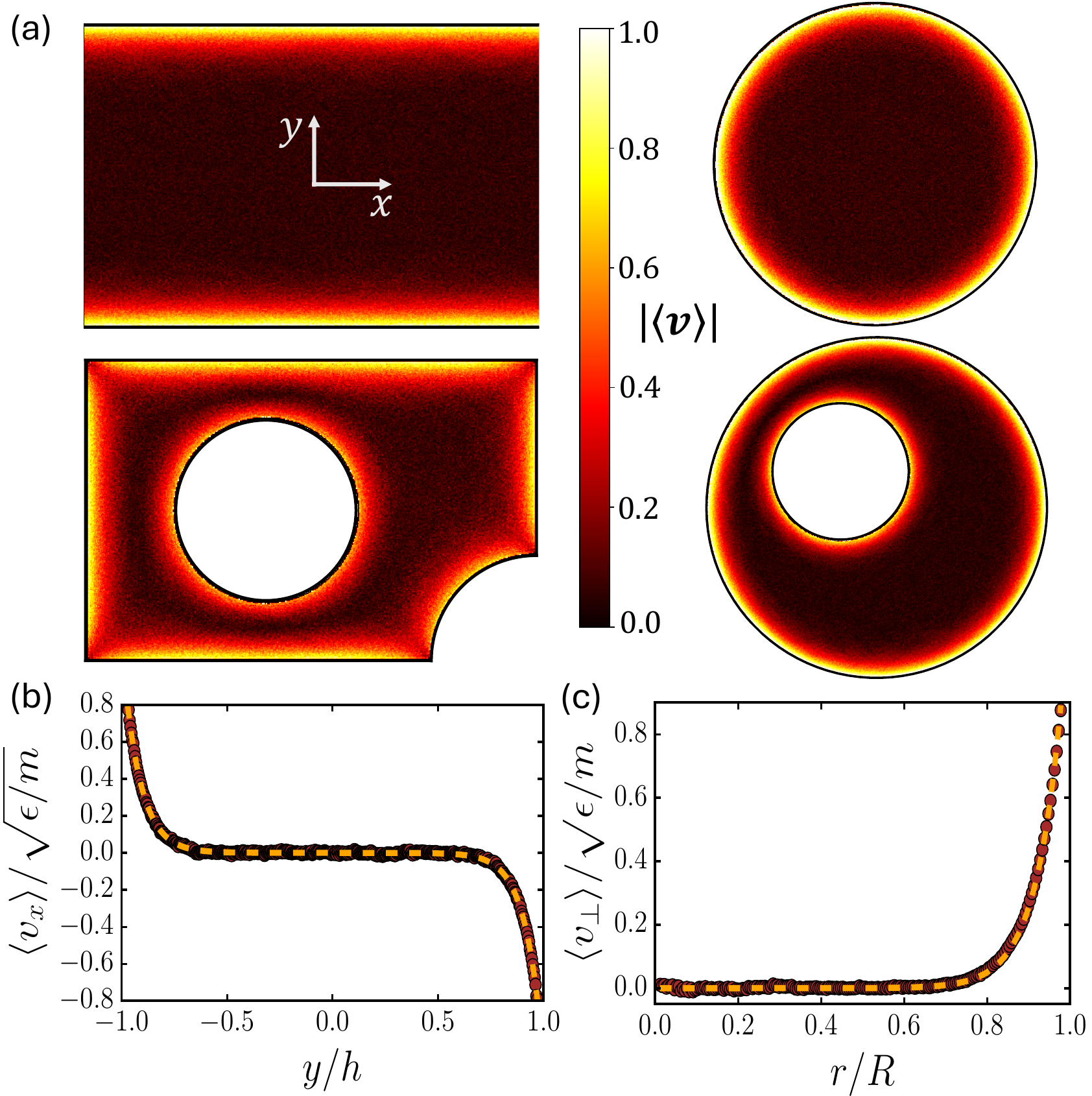}
    \caption{(a) Universality of edge currents. The unidirectional edge current persists over obstacles. The color scale shows the scaled absolute value of the averaged velocity field. Black solid lines represent the walls. Localized velocity profiles for (b) the parallel-plate geometry [top left] with side lengths $L_x=2L_y=300\sigma$, where $h=L_y/2$ and (c) the circular geometry [top right] with radius $R=100 \sigma$  for $\tau_a=2.5 \epsilon$, $\rho=0.8 m\sigma^{-2}$ and $\gamma=0.1\sqrt{m\epsilon\sigma^{-2}}$. Dashed lines are the hydrodynamic profiles computed as described in the text.}
    \label{fig:generic_boundaries}
\end{figure}

Across all of the geometries we considered, the chiral active liquid spontaneously developed a unidirectional persistent edge current localized along the external boundary, whose chirality matches that of the local active torque. Along internal boundaries, the chirality of the edge current is opposite to the locally applied torque. Figure \ref{fig:generic_boundaries}(a) shows some examples of generic boundaries, where the size of these confinements is two orders of magnitude larger than the size of a single dimer.

From these microscopic equations, we can derive hydrodynamic equations and the procedure is summarized in the SM \cite{dean1996langevin, nakamura2009derivation,Irving1950}. The resulting noise-averaged equation in the incompressible limit for the velocity field $\mathbf{v}(\mathbf{r},t)$ is
\begin{equation}
        \rho D_t \textbf{v}=-\frac{\gamma}{m}\rho \textbf{v}-\nabla p +(\eta_\mathrm{S}+\eta_\mathrm{R})\nabla^2 \textbf{v} +\eta_\mathrm{O}\nabla^2 \textbf{v}^*,
\end{equation}
where $D_t =(\partial_t+\textbf{v}\cdot \nabla)$ is the convective derivative, and incompressibility requires  $\nabla \cdot \textbf{v}=0$. The density $\rho$ is assumed constant, and $p$ is the pressure, $\eta_\mathrm{S},\eta_\mathrm{R}$ and $\eta_\mathrm{O}$ are the shear, rotational and odd viscosities, while $\textbf{v}^*$ is $\textbf{v}$ rotated $90^\circ$ clockwise. We consider these equations in two geometries, a parallel-plate confinement and a circular confinement. In the parallel-plate confinement, the system is periodic along the $x$-axis with length $L_x$ and bounded by two parallel walls at  $y=\pm L_y/2$. In the circular confinement, we consider a circular wall of radius $R$.  If an edge current is assumed, the hydrodynamic equations support an exponentially localized velocity profile tangential to the wall~\cite{Poggioli2023_Kelvin}. While the full solutions of the velocity profile depend on the global geometry, the tangential velocity close to the wall is
\begin{equation}
    \begin{aligned}
        v(b)& \sim V \exp[-\kappa_\gamma b],
    \end{aligned}
\end{equation}
where $\kappa_\gamma=\sqrt{\gamma \rho/m(\eta_S+\eta_R)}$,  $b$ is the distance away from the wall within a chosen hydrodynamic region \cite{Chen2015,Poggioli2023_Kelvin}, and $V$ is the amplitude. This form is the result of a balance between the substrate drag and the shear stress, supplied with a boundary condition $v(b\to \infty)\to 0$.  The amplitude cannot be determined via boundary conditions, but is given by the average orbital angular momentum of the system. 

Given the exponentially localized edge velocity profile, the hydrodynamic equations are linear and allow for superposition across multiple boundaries independently. Each boundary supports its own localized mode, provided their separation is much larger than $\kappa^{-1}_\gamma$. This superposition principle explains the universality shown in Fig. \ref{fig:generic_boundaries}(a). The viscosities of this system can be measured by driving the system with a known external force field in periodic conditions and measuring the resulting steady-state velocity and spin profiles \cite{hess2002determining}. We find that $\eta_S\gg\eta_R$ for the conditions we study, while $\eta_O$ does not enter our calculations, so we do not investigate it. The derivations and full form of the edge velocity profiles and the values of viscosities can be found in the SM. Using these independently evaluated transport coefficients, Fig.~\ref{fig:generic_boundaries}(b) and (c) show that we are able to  match the thickness of the edge current with this hydrodynamic theory. While the hydrodynamic equations can predict the localization profile, they are unable to deduce its magnitude or connect it to the microscopic parameters. Moreover, the sound-wave dispersion relation for this system is gapless~\cite{Poggioli2023_Kelvin}, obscuring topological arguments from explaining the origin and robustness of these edge currents. However, the connection between edge currents and microscopic details is manifest if we consider the total angular momentum of the system.

The total angular momentum of the system can be expressed as the sum of the total orbital $L$ and spin $S$ angular momenta
\begin{equation}
    J=L+S=\sum_{i\alpha} \textbf{r}_{i\alpha}\times \textbf{p}_{i\alpha}.
\end{equation}
where $L=\sum_i (\textbf{r}_{i1}+\textbf{r}_{i2})\times (\textbf{p}_{i1}+\textbf{p}_{i2})/2$ and $S=\sum_i (\textbf{r}_{i1}-\textbf{r}_{i2})\times (\textbf{p}_{i1}-\textbf{p}_{i2})/2$. The equation of motion describing the evolution of the total angular momentum density $j=J/A$, where $A$ is the area, can be derived from the underlying Langevin equations,
\begin{equation}\label{eq:djdt}
    \frac{dj}{dt}=-\frac{\gamma}{m}j+\frac{\rho}{2m} \tau_a +\frac{1}{A}\sum_{i\alpha} \textbf{r}_{i\alpha}\times \textbf{F}_{i\alpha}^{\text{w}}+\sqrt{2\gamma \kB T \frac{\mathcal{I}_m}{mA^2}}\zeta(t),
\end{equation}
where $\rho=2mN/A$ is the bulk mass density, $\mathcal{I}_m=\sum_{i\alpha} m\expval{r^2_{i\alpha}}$ is the noise-averaged moment of inertia of the system, and $\zeta(t)$ is a Gaussian white noise with zero mean and unit variance. In the SM, we show that in frictionless confinements with slowly varying curvature, the mean contribution from the wall forces vanishes. Walls that impose frictional forces are additionally considered in the SM but do not qualitatively change the system's behavior.

\begin{figure}
    \centering
    \includegraphics[width=0.9\columnwidth]{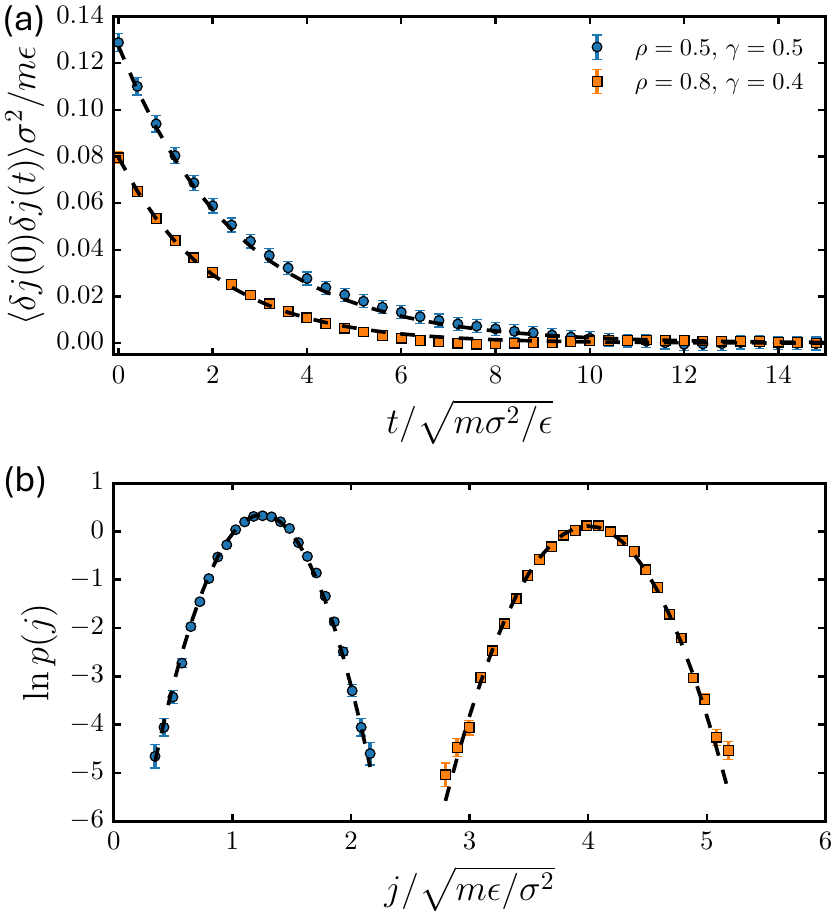}
    \caption{Statistics of the total angular momentum density. (a) The autocorrelation function of the total angular momentum density in the circular confinement. (b) Distribution of the angular momentum density. Dashed lines are the theoretical predictions. }
    \label{fig:angular momentum statistics}
\end{figure}

In the case of circular confinement, the frictionless wall contribution vanishes identically, and the resulting stochastic equation~\ref{eq:djdt} is an Ornstein-Uhlenbeck process. Its linearity allows us to deduce the time-correlation function of the angular momentum density in steady state, which is exponentially decaying $\expval{\delta j(0)\delta j(t)} = \expval{\delta j^2}\mathrm{exp}[-\gamma t/m]$. This is shown in Fig. \ref{fig:angular momentum statistics}(a) to be in excellent agreement with that computed directly from simulations. Moreover, we can solve for the steady-state distribution $P(j)$ given by

\begin{equation}
\begin{aligned}
    P(j)= \frac{1}{\sqrt{2\pi \expval{\delta j^2}}} &\exp\qty[- \frac{(j-\expval{j})^2}{2\expval{\delta j^2}}], \\
\end{aligned}
\end{equation}

\noindent which is Gaussian with a mean $\expval{j}=\rho \tau_a/2\gamma $ that is a linear function of the active torque, with a constant of proportionality given by the density and substrate friction. The variance is $\expval{\delta j^2}=\rho \kB T/2\pi$. We note here the simplicity of the statistics of the total angular momentum, which are Gaussian and thermal. The only contribution from activity is a shift in the mean value. Figure \ref{fig:angular momentum statistics}(b) shows the numerical results along with the theoretical predictions for the total angular momentum statistics in circular confinements, demonstrating good agreement.

\begin{figure}
    \centering
    \includegraphics[width=0.85\columnwidth]{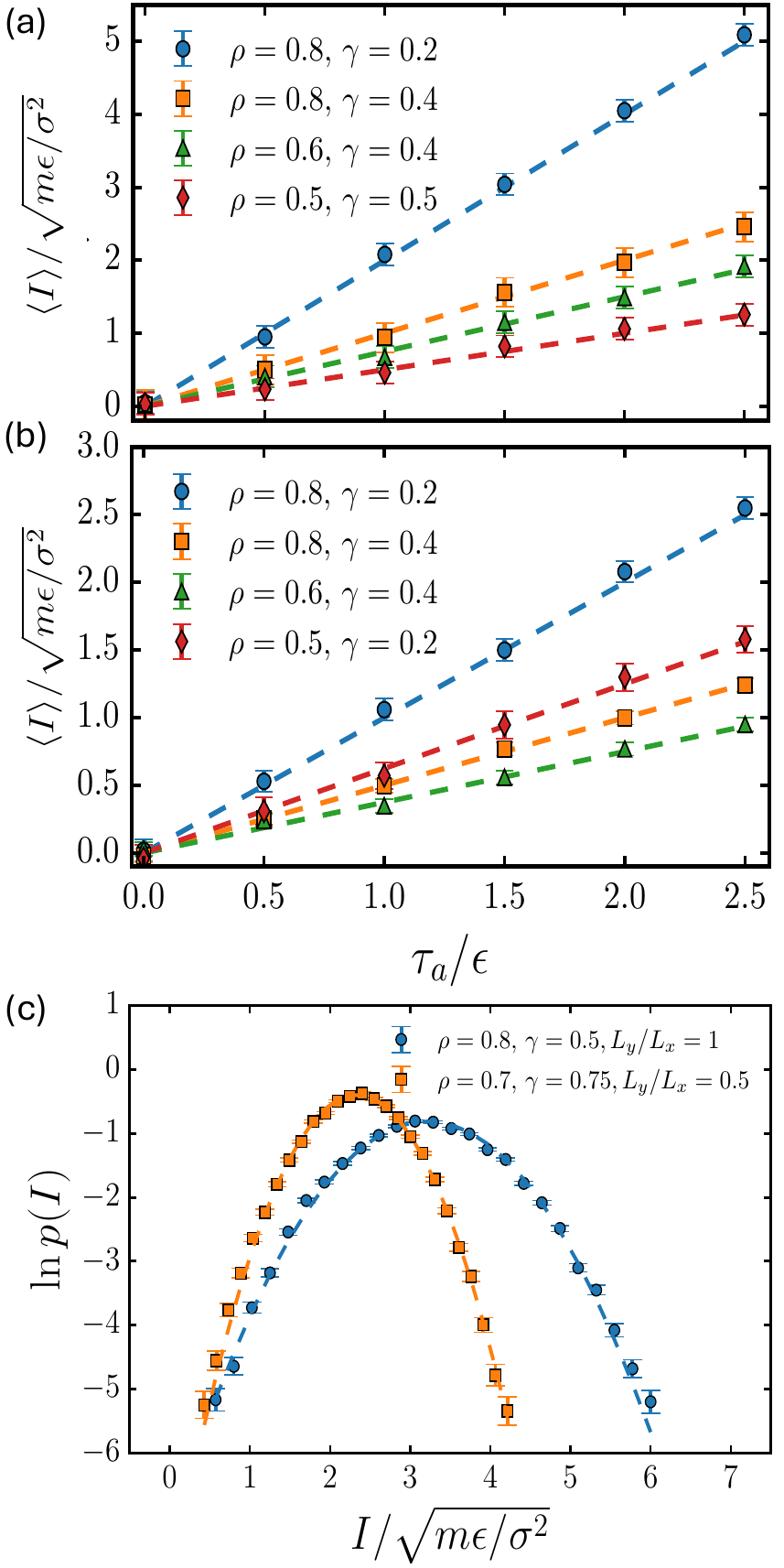}
    \caption{Edge current as a function of active torque for  parallel-plate, $L_x=L_y=400\sigma$ (a) and circular, $R=250\sigma$ (b) geometries at different densities and drag coefficients. The dashed lines are the predictions from the balance of angular momentum. (c) Steady-state edge current distribution for the parallel-plate confinement for different parameters.
}
    \label{fig:current vs tau and velocity profiles}
\end{figure}

The angular momentum of this system can be stored as either spin or orbital angular momenta. In the dilute limit, where collisions between dimers are rare, approximately all of the injected angular momentum is stored in the spin degrees of freedom and dissipated through them, with a spin steady-state average value of $\expval{S}/A\approx \rho\tau_a/2\gamma$. However, as the system's density increases, collisions between the dimers act to transfer spin angular momentum into orbital angular momentum. Numerically, we find that $\expval{L}/ \expval{S} \sim \mathcal{O}(100)$ in dense systems $\rho\sigma^2/m \gtrsim 0.5$ (SM). 

In the limit of high density, the dimers' spin becomes frustrated and unable to build up, transferring approximately all of the injected angular momentum into orbital angular momentum. The steady-state orbital angular momentum can be written as
\begin{equation}
    \expval{L}=\int \textbf{r}\times \rho\textbf{v}(\textbf{r})  \;dA,
\end{equation}
where $\rho\textbf{v}(\textbf{r})$ is the mean momentum field and $dA$ is an area element with an integral covering the domain. Since our model is dissipative, isotropic and homogeneous, there cannot be a finite average flow field in the bulk of the system, since gradients leading to a net flow vanish. However, the homogeneity breaks down close to the boundary of the system. Therefore, a mean velocity field can be concentrated close to the boundary
\begin{equation}
\begin{aligned}
    \expval{L}&\approx  \int \textbf{r}_\text{e}(l)\times \hat{\mathbf{e}}_t  \;\rho v(b)  \;dl \; db,\\
\end{aligned}
\end{equation}
where $\textbf{r}_\text{e}$ is the position vector of a point along the boundary of the system, parameterized by the length along the boundary $l$, $\textbf{v}=v \hat{\mathbf{e}}_t$ is the edge velocity pointing along the tangential direction to the wall $\hat{\mathbf{e}}_t$, whose magnitude $v$ is assumed to depend only on the normal distance $b$ away from the boundary. The density oscillations near the boundary are also assumed to depend on $b$ alone. These assumptions are valid in the limit of small curvatures, where the local geometry of the boundary is flat. This approximation allows us to factorize the integral into  $\expval{L}= 2\expval{I} A$, where the mean current in the continuum limit is
\begin{equation}
    \expval{I}=\int \rho(b) v(b)\;db.
\end{equation}
Therefore, in steady state, the mean current in a dense system is directly proportional to the total mean angular momentum density 
\begin{equation}\label{eq:Ivtau}
    \expval{I}=\frac{\expval{j}}{2}=\frac{\rho}{4\gamma} \tau_a.
\end{equation}
This result is akin to Ohm's law as derived through the Drude model \cite{simon2013oxford}, where the current is linearly proportional to the driving force, with a proportionality constant depending on the density and drag coefficient. 

Equation \ref{eq:Ivtau} is remarkable as it links the locally applied active torque to the globally driven edge current through a response relation that is valid far from equilibrium.  Figures \ref{fig:current vs tau and velocity profiles}(a) and (b) show the mean edge current measured in simulations across a large range of  parameters, showing good agreement with our predictions. This result is accurate as long as the localization assumption is valid, since the first-order correction is proportional to $(\kappa_\gamma  a)^{-1}$, where $a$ is the linear size of the confinement. Note that when the edge current is accurately described by the continuum theory, $\sigma \ll \kappa_\gamma^{-1}$, the amplitude of the velocity profile can then be given as $V=\kappa_\gamma \expval{I}/\rho$, which resolves the ambiguity from the hydrodynamic equations.
An analogous analysis applies to systems where the spin is not quenched  \cite{yang2020robust,soni2019odd,liu2020oscillating} as well as systems where there is no explicit spin degree of freedom \cite{metzger2026equation,Langford2025} (see SM). 

The fluctuations in the edge current are related to those of the total angular momentum, but not equivalent as in the case of the mean value. Unlike the mean, which is localized at the boundary, fluctuations in the angular momentum are distributed throughout the bulk, so the approximation connecting the mean current to the mean angular momentum is not valid for their fluctuations. The instantaneous edge current is defined in circular confinement as $I=\frac{1}{2\pi R}\sum_{i\alpha} \textbf{p}_{i\alpha}\cdot \hat{e}_{t,i\alpha}$, and in parallel-plate confinement as $I=- \frac{1}{L_x}\sum_{i\alpha} \textbf{p}_{i\alpha}\cdot \hat{e}_x  \; y_{i\alpha}/\abs{y_{i\alpha}}  $, where $\hat{e}_{t,i\alpha}$ is the tangential unit vector at the location of the $i\alpha$'th particle, $\hat{e}_x$ is the unit vector along the $x$-axis, and $y_{i\alpha}/\abs{y_{i\alpha}}$ is an indicator function for whether the particle is in the upper or lower half. From these microscopic definitions, we can derive governing stochastic equations for the edge currents, as detailed in the SM. The resulting equation is
\begin{equation}
   \frac{d}{dt}I=-\frac{\gamma}{m}I+\frac{n\rho}{4m}\tau_a +\sqrt{2\gamma \kB T \frac{\rho}{m}\mathrm{G}} \;\zeta(t),
\end{equation}
where $n=\{1,2\}$ and $\mathrm{G}=\{1/4\pi, L_y/L_x\}$ for the circular and parallel-plate confinements, respectively. This equation supports a steady-state distribution of the form
\begin{equation}
\begin{aligned}
     P(I)= \frac{1}{\sqrt{2\pi \expval{\delta I^2}}} \exp&\qty[-\frac{(I-\expval{I})^2}{2\expval{\delta I^2}}],\\
\end{aligned}
\end{equation}
where $\expval{\delta I^2} =\rho \kB T\mathrm{G}$. Figure \ref{fig:current vs tau and velocity profiles}(c) shows the steady state distribution of edge currents, again with quantitative agreement with our theoretical predictions. From Gallavotti-Cohen symmetry \cite{gallavotti1995dynamical}, this implies that the rate of entropy production, $\dot{\Sigma}$, can be inferred directly from the edge current, $\dot{\Sigma}=I \tau_a/2 \kB T \gamma G$. The Gaussian form is a consequence of the edge current being a free variable, stemming from the structure of the global balance of angular momentum. The form of its mean and variance is set by the system in question.

In this letter, we elucidated the physical origin of edge currents in chiral active liquids by revealing a direct relationship between the global injection of angular momentum and its dissipation through the spontaneous edge currents. This approach resulted in a clean connection between the mean edge current and activity, expressed as an Ohmic-like conductance law. This simple result reveals an unusual physical picture, where a locally applied driving force leads to an intensive global current in the form of a linear response that is valid arbitrarily far from equilibrium. Moreover, we showed that the statistics of the total angular momentum are closely related in character to those of the edge current. In both cases, we saw that the nonequilibrium steady-state distribution for this system is a shifted equilibrium distribution. Arriving at these seemingly simple results starting from the fluctuating hydrodynamic equations would be a formidable task. We validated our predictions using molecular dynamics simulations of large systems of chiral active dimers. Our work paves the way for using global balance laws as a starting point for analyzing collective behavior in systems driven far from equilibrium.

\bigskip

\textbf{Acknowledgments}
This work was supported by the U.S. Department of Energy, Office of Science, Office of Advanced Scientific Computing Research, and Office of Basic Energy Sciences, via the Scientific Discovery through Advanced Computing (SciDAC) program.

\bibliography{references}

\end{document}


\title{Supplemental Material for ``Origin of Edge Currents in Chiral Active Liquids"}
\author{Faisal Alsallom}
\email{alsallom@berkeley.edu}
\affiliation{Department of Physics, University of California, Berkeley, California 94720, USA}
\affiliation{Chemical Science Division, Lawrence Berkeley National Laboratory, Berkeley, California 94720, USA}

\author{David T. Limmer}
\email{dlimmer@berkeley.edu}
\affiliation{Department of Chemistry, University of California, Berkeley, California 94720, USA}
\affiliation{Materials Science Division, Lawrence Berkeley National Laboratory, Berkeley, California 94720, USA}
\affiliation{Chemical Science Division, Lawrence Berkeley National Laboratory, Berkeley, California 94720, USA}
\affiliation{Kavli Energy NanoSciences Institute, Berkeley, California 94720, USA}

\date{\today}

\maketitle

\section{Simulation details}
We consider a system composed of $N$ identical dimers in two dimensions. Every dimer consists of two equivalent monomers, each of mass $m$. We evolve our system according to the under-damped Langevin equations of motion
\begin{equation}\label{Dumbbell Model}
    \begin{aligned}
        \dot{\textbf{r}}_{i\alpha}&=\frac{\textbf{p}_{i\alpha}}{m}\\
        \dot{\textbf{p}}_{i\alpha}&=-\frac{\gamma}{m}\textbf{p}_{i\alpha}+\textbf{F}_{i\alpha}^c+\textbf{F}_{i\alpha}^a +\sqrt{2\gamma \kB T} \boldsymbol{\eta}_{i\alpha}(t)
    \end{aligned}
\end{equation}
where $i\in\{1,\dots,N\}$ indexes the dimer and $\alpha \in\{1,2\}$ indexes the monomer. $\textbf{r}_{i\alpha}$ and $\textbf{p}_{i\alpha}$ are the position and momentum vectors of the $i\alpha$th particle, respectively. $\textbf{F}_{i\alpha}^c$ is the total force due to potential forces acting on the $i\alpha$th particle, and $\textbf{F}_{i\alpha}^a$ is the active force acting on the $i\alpha$th particle. Within each dimer, each monomer experiences equal and opposite active forces, such that the forces are perpendicular to the instantaneous bonding vector $\textbf{d}_i$. The active torque $\tau_a$ on each dimer is constant and equal, $F_{i\alpha}^a(t) d_i(t)=\tau_a$. $\gamma$ is the friction coefficient of the underlying bath, and $T$ is its temperature. $\boldsymbol{\eta}_{i\alpha}$ is a Gaussian white random noise, with statistics
\begin{equation}
    \expval{\boldsymbol{\eta}_{i\alpha}}=0, \hspace{0.3cm} \expval{\boldsymbol{\eta}_{i\alpha}(t) \otimes \boldsymbol{\eta}_{j\beta}(t')}=\delta_{ij}\delta_{\alpha \beta} \delta(t-t') \mathrm{I}_{2\times 2},
\end{equation}
where $\mathrm{I}_{2\times 2}$ is the identity. In this model, we treat the bath as passive and quiescent. We do not consider any underlying hydrodynamics of the bath. This model could be realized by sedimenting the chiral active particles on a solid substrate. This model would be classified as a ``dry model'' \cite{Marchetti2013}.

 The interaction potential energy between non-bonded monomers is given by the Weeks-Chandler-Andersen short-ranged repulsive interaction potential \cite{Weeks1971_WCA},
\begin{equation}
    u_{\text{WCA}}(r)= \begin{cases}
        4\epsilon \left[\left(\frac{\sigma}{r}\right)^{12}-\left(\frac{\sigma}{r}\right)^{6}\right]+\epsilon & r<2^{1/6} \sigma, \\
        0 & r>2^{1/6} \sigma,
    \end{cases}
\end{equation}
where $\sigma$ represents the particle's diameter and $\epsilon$ sets an energy scale. The bonding potential energy within a dimer is given by $k(\abs{\textbf{d}_i}-\sigma)^2 /2$, where $\textbf{d}_i$ is the relative position vector connecting the two bonded monomers.
We simulate this system using molecular dynamics simulations in both periodic and confined boundary conditions. In the confined boundary conditions, we model the confining wall as a frictionless hard wall that interacts with repulsive Weeks-Chandler-Andersen potential with $\epsilon_\text{wall}=10 \epsilon$ and $\sigma_\text{wall}=\sigma$. In our simulations, we use the following parameters: $k=200 \epsilon \sigma^{-2}$, $\epsilon=\sigma=T=1$. The values of $\gamma, \tau_a$ and $N$ vary, as well as the system sizes considered.  $N$ ranges from $10^4$ to $  10^5$. For thermalization runs, we used time steps $\Delta t=0.001\sqrt{m\sigma^2/\epsilon}$, while for production runs we used $\Delta t$ up to $0.0025 \sqrt{m\sigma^2/\epsilon}$. Simulations were run for times between $5\times10^3 \sqrt{m\sigma^2/\epsilon}$ and $1\times10^4 \sqrt{m\sigma^2/\epsilon}$ to converge the statistics shown in the main text.

\section{Hydrodynamics}
We summarize the procedure of deriving the hydrodynamic equations in our system following the approach in \cite{Poggioli2023_Kelvin}.  The hydrodynamic equations are conveniently expressed through 
\begin{equation}
    \begin{aligned}
        \textbf{R}_{i}&\equiv \frac{\textbf{r}_{i1}+\textbf{r}_{i2}}{2},\\
        \textbf{P}_i&\equiv \textbf{p}_{i1}+\textbf{p}_{i2},\\
        s_i &\equiv \frac{1}{2}(\textbf{r}_{i1}-\textbf{r}_{i2})\times (\textbf{p}_{i1}-\textbf{p}_{i2}),
    \end{aligned}
\end{equation}
where $\textbf{R}_{i}$ is  the center of mass of dimer $i$, with center of mass momentum $\textbf{P}_i$ and $s_i$ the local spin. The time evolution of these coordinates is derived from the monomer equations of motion as
\begin{equation}
    \begin{aligned}
        \dot{\textbf{R}}_i&=\frac{\textbf{P}_i}{2m},\\
        \dot{\textbf{P}}_i&=-\frac{\gamma}{m}\textbf{P}_i + \textbf{F}_i^c+\sqrt{4\gamma \kB T} \boldsymbol{\eta}_i (t),\\
        \dot{s}_i&=-\frac{\gamma}{m}s_i +\frac{1}{2}\textbf{d}_i \times \Delta \textbf{F}_i^c+\tau_a +\sqrt{\gamma \kB T d^2} \mu_i(t),
    \end{aligned}
\end{equation}
where $\boldsymbol{\eta}_i$ and $\mu_i$ are delta correlated, zero mean, unit variance Gaussian random variables,  $ \textbf{F}_i^c = \textbf{F}_{i1}^c +\textbf{F}_{i2}^c$  and $\Delta \textbf{F}_i^c= \textbf{F}_{i1}^c -\textbf{F}_{i2}^c$. We approximate the length of the bond to be constant, $d$, in writing the spin equation of motion, which is a good approximation given the stiffness of the spring constant used. 

With these definitions, we may define the hydrodynamic fields for mass density, linear momentum density and spin angular momentum density
\begin{equation}
    \begin{aligned}
        \rho(\textbf{x},t)&=\sum_i 2m \;\delta(\textbf{x}-\textbf{R}_i(t)),\\
        \rho \textbf{v}(\textbf{x},t)&=\sum_i \textbf{P}_i(t) \;\delta(\textbf{x}-\textbf{R}_i(t)),\\
        \rho\mathcal{S}(\textbf{x},t)&=\sum_i s_i(t)\; \delta(\textbf{x}-\textbf{R}_i(t)),
    \end{aligned}
\end{equation}
with each sum over all $N$ dimers.  The equations of motion for these fields can be derived through the equations of motion of the microscopic degrees of freedom. The resulting equation of motion for the density is
\begin{equation}
        \partial_t \rho +\nabla \cdot (\rho \textbf{v})=0,
 \end{equation}
 that for the linear momentum density is
\begin{equation}
    \begin{aligned}
        \partial_t(\rho \textbf{v})+\nabla \cdot(\rho \textbf{v}\otimes \textbf{v})&= -\frac{\gamma}{m}\rho \textbf{v}+ \sum_{i} \textbf{F}_i^c \delta(\textbf{x}-\textbf{R}_i)\\
        + &\sqrt{4\gamma \kB T} \sum_i \boldsymbol{\eta}_i(t) \delta(\textbf{x}-\textbf{R}_i),
    \end{aligned}
\end{equation}  
    and finally for the spin angular momentum density,
\begin{equation}
    \begin{aligned}      
        \partial_t (\rho \mathcal{S})+\nabla \cdot (\textbf{v}\; \rho \mathcal{S})&=-\frac{\gamma}{m} \rho \mathcal{S}+\frac{\tau_a}{2m} \rho\\
        +&\frac{1}{2}\sum_i \textbf{d}_i \times \Delta \textbf{F}_i^c \delta(\textbf{x}-\textbf{R}_i)\\
         +&\sqrt{\gamma \kB T d^2}\sum_i \mu_i(t) \delta(\textbf{x}-\textbf{R}_i),
    \end{aligned}
\end{equation}
The sum over the noises can be evaluated through the central limit theorem. In the case of the momentum balance equations,  the mean is
\begin{equation}
    \left\langle \sqrt{4\gamma \kB T} \sum_i  \boldsymbol{\eta}_i \delta(\textbf{x}-\textbf{r}_i) \right\rangle =0,
\end{equation}
while for the variance
\begin{equation}
\begin{aligned}
    & \left \langle 4\gamma \kB T \sum_i \sum_j \boldsymbol{\eta}_i \otimes \boldsymbol{\eta}_j \delta(\textbf{x}-\textbf{r}_i) \delta(\textbf{x}'-\textbf{r}_j)\right \rangle \\&= 4\gamma \kB T \sum_i \sum_j    \left\langle \boldsymbol{\eta}_i(t) \otimes \boldsymbol{\eta}_j(t') \right\rangle   \left\langle  \delta(\textbf{x}-\textbf{r}_i) \delta(\textbf{x}'-\textbf{r}_j) \right\rangle\\
     & = 4\gamma \kB T \sum_i \sum_j \mathbf{1} \delta_{ij} \delta(t-t') \left\langle  \delta(\textbf{x}-\textbf{r}_i) \delta(\textbf{x}'-\textbf{r}_j) \right\rangle\\
     & =  4\gamma \kB T \sum_i  \mathbf{1}  \delta(t-t') \delta(\textbf{x}-\textbf{x}') \left\langle   \delta(\textbf{x}'-\textbf{r}_i) \right\rangle\\
     & = \frac{2\gamma \kB T  \rho(\mathbf{x},t)}{m } \mathbf{1} \delta(t-t') \delta(\textbf{x}-\textbf{x}').
\end{aligned}
\end{equation}
Therefore, we can rewrite the sum over the microscopic noises as
\begin{equation}
    \sqrt{4\gamma\kB T} \sum_i  \boldsymbol{\eta}_i \delta(\textbf{x}-\textbf{r}_i) = \sqrt{\frac{2\gamma \kB T \rho(\mathbf{x},t)}{m}} \boldsymbol{\xi}(\textbf{x},t),
\end{equation}
where $\boldsymbol{\xi}(\textbf{x},t)$ is a space-time delta-correlated Gaussian noise. A similar procedure can be followed for the noise contributions to the spin balance equation.

The conservative forces term in the momentum equation can be written in the following way
\begin{equation}
\begin{aligned}
    \sum_i &\textbf{F}_{c,i} \delta(\textbf{x}-\textbf{r}_i) = \sum_{i,j}  \textbf{F}_{c,j \text{ on } i} \delta(\textbf{x}-\textbf{r}_i)\\
    & = \frac{1}{2} \sum_{i,j} \left[ \textbf{F}_{c,j \text{ on } i} \delta(\textbf{x}-\textbf{r}_i) - \textbf{F}_{c,i \text{ on } j} \delta(\textbf{x}-\textbf{r}_i)\right]\\
    & = \frac{1}{2} \sum_{i,j} \left[ \textbf{F}_{c,j \text{ on } i} \delta(\textbf{x}-\textbf{r}_i) - \textbf{F}_{c,j \text{ on } i} \delta(\textbf{x}-\textbf{r}_j)\right]\\
    & = \frac{1}{2} \sum_{i,j} \textbf{F}_{c,j \text{ on } i} \left[ \delta(\textbf{x}-\textbf{r}_i) - \delta(\textbf{x}-\textbf{r}_i+ (\textbf{r}_i-\textbf{r}_j)) \right]\\
    & = - \frac{1}{2} \sum_{i,j}   \textbf{F}_{c,j \text{ on } i}  \otimes (\textbf{r}_i-\textbf{r}_j)\cdot \partial_{\textbf{x}} \delta(\textbf{x}-\textbf{r}_i )\\
    & = \nabla \cdot \left [-\frac{1}{2} \sum_{i,j} \textbf{F}_{c,j \text{ on } i} \otimes \textbf{r}_{ij}  \delta(\textbf{x}-\textbf{r}_i) \right ],
\end{aligned}
\end{equation}
through which we define the Irving-Kirkwood virial stress \cite{Irving1950} as
\begin{equation}
    \tau^V(\mathbf{x},t) = -\frac{1}{2} \sum_{i,j}  \textbf{F}_{c,j \text{ on } i}\otimes  \textbf{r}_{ij} \delta(\textbf{x}-\textbf{r}_i).
\end{equation}
This rewriting leads to the following form of the momentum balance equation
\begin{equation}
   \partial_t(\rho \textbf{v}) + \nabla \cdot (\textbf{v} \otimes \rho \textbf{v}) =  - \frac{\gamma}{m} \rho \textbf{v} + \nabla\cdot \tau^V +\sqrt{\frac{2\gamma \rho}{m\beta}} \boldsymbol{\xi}(\textbf{x},t).
\end{equation}
This equation, along with the spin balance equation, is further developed in \cite{Poggioli2023_Kelvin} by averaging over noise and relating the stress tensors to the kinematic fields  phenomenologically via transport coefficients as allowed by symmetry. This process yields the following set of hydrodynamic equations in the incompressible limit
\begin{equation}
    \begin{aligned}
        \nabla \cdot \textbf{v}&=0,\\
        \rho D_t \textbf{v}&=-\frac{\gamma}{m}\rho \textbf{v}-\nabla p +(\eta_\mathrm{S}+\eta_\mathrm{R})\nabla^2 \textbf{v} +\eta_\mathrm{O}\nabla^2 \textbf{v}^*,\\
        \rho D_t \Delta \mathcal{S}&=-\frac{\gamma}{m}\rho\Delta \mathcal{S}+2\eta_\mathrm{R} \omega,
    \end{aligned}
\end{equation}
where $D_t=\partial_t+\textbf{v}\cdot \nabla$ is the material derivative,  $p$ is the pressure, $\eta_\mathrm{S},\eta_\mathrm{R}$ and $\eta_\mathrm{O}$ are the shear, rotational and odd viscosities, and $\textbf{v}^*$ is $\textbf{v}$ rotated $90^\circ$ clockwise, $\omega=\nabla \times \textbf{v}$ is the vorticity and $\Delta \mathcal{S}$ is the spin density relative to the bulk steady-state value. 

While the viscosities are phenomenological, they can be measured numerically by driving the system with a known force field. In our simulations, we follow \cite{Poggioli2023_Kelvin} in applying a sinusoidal force on the monomers in periodic boundary conditions of the form $\textbf{F}_d(x)=F_o \sin(ky) \hat{e}_x$, such that $k$ is small compared to $\sigma^{-1}$ and matches the boundary conditions. In steady-state, the resulting flow profile is
\begin{equation}
    \textbf{v}=\frac{F_o}{\gamma[1+(k/\kappa_\gamma)^2]}\sin(ky)\hat{e}_x,
\end{equation}
where $\kappa_\gamma=\sqrt{\gamma \rho/m(\eta_\mathrm{S} +\eta_\mathrm{R})}$ . By measuring the amplitude of the shear flow, we can numerically deduce $\eta_\mathrm{S}+\eta_\mathrm{R}$. Similar measurements for the spin and pressure profiles can determine the viscosities uniquely. We expect the measured viscosities to match those of the unforced active system in the limit of small $F_o$ and $k$. For the conditions considered in Fig. 1(c) in the main text, we find  $\eta_\mathrm{S}=3.04 \sqrt{m\epsilon /\sigma^2}$ and $\eta_\mathrm{R}=3\times 10^{-3}\sqrt{m\epsilon /\sigma^2}$.

\begin{figure}
    \centering
    \includegraphics[width=0.9\columnwidth]{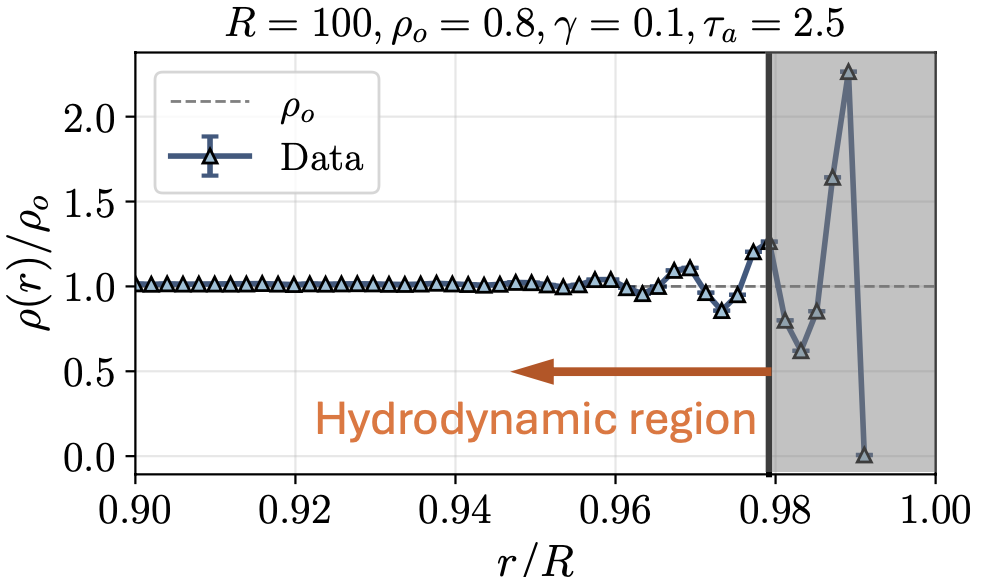}
    \caption{Density profile close to the wall in the circular confinement. Correlations induced by the wall are negligible within the hydrodynamic region.}
    \label{fig:hydrodynamic_region}
\end{figure}

It is important to note that these hydrodynamic equations are only valid in the bulk. The walls introduce correlations in density and momentum fields that are not accounted for, and as such we can only apply the hydrodynamic equations to within a hydrodynamic region, which is chosen to coincide with the second density peak away from the wall \cite{Chen2015}, see Fig.~\ref{fig:hydrodynamic_region}.

\subsection{Solutions for simple geometries}

The incompressible hydrodynamic equations can be solved in steady-state for simple geometries. We consider two examples: circular confinement, and two parallel plates confinement.

\textbf{Parallel-plate confinement.} We anticipate a steady state velocity profile of the form $\textbf{v}=v_x(y)\hat{e}_x$  with one boundary condition $v_x(y=0)=0$ due to symmetry and a pressure profile $p=p(y)$. The steady-state momentum balance equation along the $x$-axis is
\begin{equation}
    0=-\kappa_\gamma^2  v_x+\partial_y^2 v_x.
\end{equation}
where $\kappa_\gamma^2 = \gamma \rho /m (\eta_\mathrm{S}+\eta_\mathrm{R})$. The solution is conveniently expressed as
\begin{equation}
    v_x(y)=-V \frac{\sinh(\kappa_\gamma y)}{\sinh(\kappa_\gamma h_\text{hyd})}
\end{equation}
where $h=L_y/2$ and $V$ is a constant determined via the average orbital angular momentum of the system
\begin{equation}
    \expval{L}_\text{hyd}=\int_{-h_\text{hyd}}^{h_\text{hyd}} y\; \rho L_xv_x(y) dy.
\end{equation}
This average angular momentum within the hydrodynamic region can be measured numerically in order to obtain a numerical value for the amplitude $V$.

\textbf{Circular confinement.} We anticipate a steady-state velocity profile of the form $\textbf{v}= v_\perp (r)\hat{\mathbf{e}}_t$, with one boundary condition $v_\perp(r=0)=0$ and a pressure profile $p=p(r)$. The steady-state momentum balance equation along the polar direction is
\begin{equation}
    0=-\kappa_\gamma^2 v_\perp+\qty(\partial_r^2+\frac{1}{r}\partial_r-\frac{1}{r^2})v_\perp \,
\end{equation}
whose solution is given by
\begin{equation}
v_\perp(r) = V \frac{J_1(i \kappa_\gamma r)}{J_1(i \kappa_\gamma R_\text{hyd})}
\end{equation}
where $J_1$ is the first order Bessel function of the first kind. In the limit of large circular confinements, the tangential velocity profile is well-approximated by
\begin{equation}
    v_\perp(r)=V \exp\qty(-\kappa_\gamma (R_\text{hyd}-r)),
\end{equation}
where $V$ is a constant determined via the average orbital angular momentum
\begin{equation}
     \expval{L}_\text{hyd}=\int_{0}^{R_\text{hyd}} r \cdot2\pi r\rho  v_\perp(r) dr.
\end{equation}
in analogy with the parallel plate confinement.

\section{Balance of Angular momentum}
The total angular momentum of the system with respect to some reference point can be expressed as
\begin{equation}\label{total angular momentum definition}
    J=\sum_{i\alpha} \textbf{r}_{i\alpha}\times \textbf{p}_{i\alpha},
\end{equation}
and its equation of motion can be derived from the governing under-damped Langevin equations for $\textbf{p}_{i\alpha}$
\begin{equation}
    \dot{J}=-\frac{\gamma}{m}J+N \tau_a +\sum_{i\alpha} \textbf{r}_{i\alpha}\times \qty[\textbf{F}_{i\alpha}^{\text{w}}+\sqrt{2\gamma \kB T}\boldsymbol{\eta}_{i\alpha}],
\end{equation}
where the contribution from the central forces between the monomers vanishes due to Newton's third law, and $\textbf{F}_{i\alpha}^{\text{w}}$ is the force exerted by the confining wall on the $i\alpha$th particle. The noise term can be further simplified given that positions and noise forces are uncorrelated. The mean of the sum is zero,
\begin{equation}
\begin{aligned}
    \sqrt{2\gamma \kB T} \expval{  \sum_{i\alpha} \textbf{r}_{i\alpha}\times \boldsymbol{\eta}_{i\alpha}}&=\sqrt{2\gamma \kB T} \sum_{i\alpha} \expval{\textbf{r}_{i\alpha}}\times \expval{\boldsymbol{\eta}_{i\alpha}}\\
    & = 0.
\end{aligned}
\end{equation}
The variance is
\begin{equation}
    2\gamma \kB T \expval{\qty(\sum_{i\alpha} \textbf{r}_{i\alpha}\times \boldsymbol{\eta}_{i\alpha}) \qty(\sum_{i\alpha} \textbf{r}_{j\beta}\times \boldsymbol{\eta}_{j\beta})}.
\end{equation}
Since $\langle \boldsymbol{\eta}_{i\alpha}(t) \otimes  \boldsymbol{\eta}_{j\beta}(t')\rangle=\delta_{ij}\delta_{\alpha \beta}\delta(t-t') \mathrm{I}_{2\times 2}$, the average can be simplified greatly by explicitly expanding the cross product in Cartesian coordinates, leading to
\begin{equation}
    2\gamma \kB T \sum_{i\alpha} \qty(\expval{x_{i\alpha}^2}+\expval{y_{i\alpha}^2}) \delta(t-t').
\end{equation}
This sum can be interpreted as the noise-averaged moment of inertia of the system, $\mathcal{I}_m$, up to a factor of $m$. We have verified that fluctuations in this quantity are minimal in the dense limit, so it can be treated as a constant.  Therefore, by the central limit theorem, the sum over the noisy torques is given by
\begin{equation}
 \sqrt{2\gamma \kB T}   \sum_{i\alpha} \textbf{r}_{i\alpha}\times \boldsymbol{\eta}_{i\alpha} =\sqrt{2\gamma \kB T \frac{\mathcal{I}_m}{m}} \zeta(t),
\end{equation}
where $\zeta$ is a random Gaussian white noise with $\expval{\zeta}=0$ and $\expval{\zeta(t)\zeta(t')}=\delta(t-t')$. 

Rewriting $N=\rho A/2m$, where $\rho$ is the bulk mass density and $A$ is the area enclosed by the confinement, we may express the equation of motion for the total angular momentum density $j\equiv J/A$ as 
\begin{equation}
    \frac{d}{dt}j=-\frac{\gamma}{m}j+\frac{\rho}{2m} \tau_a +\frac{1}{A}\sum_{i\alpha} \textbf{r}_{i\alpha}\times \textbf{F}_{i\alpha}^{\text{w}}+\sqrt{2\gamma \kB T \frac{\mathcal{I}_m}{mA^2}}\zeta(t).
\end{equation}

While the contribution from the confining wall is complicated in general, its mean behavior can be understood in the case of slowly varying curvature (compared to $\sigma$) of the wall. On average, this sum can be represented as an integral along the perimeter of the wall
\begin{equation}
    \begin{aligned}
        \sum_{i\alpha} \expval{\textbf{r}_{i\alpha}\times \textbf{F}_{i\alpha}^{\text{w}}}=\int \textbf{r}\times d\textbf{F}^\text{w},
    \end{aligned}
\end{equation}
where $\textbf{r}$ is the position vector of the wall element exerting force $d\textbf{F}^\text{w}$ on the fluid. This force element is a complicated function of the interaction potential and the density profile of the fluid. However, given the wall is frictionless, we can write $d\textbf{F}^\text{w}\equiv d\textbf{l}\times \textbf{B}(\textbf{r})$, where $d\textbf{l}$ is the length element along the wall, and $\textbf{B}(\textbf{r})=B(r)\hat{z}$ is a vector field pointing along the $z$-axis. This representation is consistent with the wall force being perpendicular at each line segment. Under the assumption of slowly varying curvature, we expect the density profile to be uniform along the perimeter of the wall, implying $B(r)=B_o$. This then reduces to the problem of the total torque exerted by a uniform, perpendicular magnetic field acting on a planar current-carrying wire, which is known to be zero.
\begin{equation}
\begin{aligned}
    \sum_{i\alpha} \expval{\textbf{r}_{i\alpha}\times \textbf{F}_{i\alpha}^{\text{w}}}&=\int \textbf{r}\times \qty(d\textbf{l}\times \hat{z})B(r)\\
    & =B_o \int \textbf{r}\times \qty(d\textbf{l}\times \hat{z})\\
    &=\textbf{0}.
\end{aligned}
\end{equation}
Therefore, the equation of motion describing the mean angular momentum density is
\begin{equation}
    \frac{d}{dt} \expval{j(t)}=-\frac{\gamma}{m} \expval{j(t)} + \frac{\rho}{2m} \tau_a,
\end{equation}
corresponding to a steady-state average value
\begin{equation}
    \expval{j}=\frac{\rho}{2\gamma}\tau_a  \, ,
\end{equation}
as shown in the main text.

\subsection{Circular confinement}

In the case of circular confinement, taking the reference point to be the center of the circle leads to a vanishing contribution from the wall, which allows us to write a simple stochastic equation of motion for the angular momentum density
\begin{equation}
    \frac{dj}{dt}=-\frac{\gamma}{m}j+\frac{\rho}{2m} \tau_a +\sqrt{\gamma \kB T \frac{\rho}{\pi m}}\zeta(t),
\end{equation}
where we substituted $\mathcal{I}_m=\rho A^2/2\pi$. This equation can be cast into a Fokker-Planck equation \cite{zwanzig2001nonequilibrium} for the probability distribution of $j$ 
\begin{equation}
    \partial_t P(j)= \frac{\gamma}{m} \partial_j \qty[(j-\expval{j}) P(j)] + \frac{\gamma \kB T\rho}{2\pi m} \partial^2_j P(j).
\end{equation}
The steady-state distribution can then be found to be a Gaussian
\begin{equation}
    P(j)= \frac{1}{\sqrt{2\pi \expval{\delta j^2}}} \exp\qty[- \frac{(j-\expval{j})^2}{2\expval{\delta j^2}}], 
\end{equation}
where $\expval{\delta j^2}=\rho \kB T/2\pi$. We note that the contribution from activity appears only as a shift to the mean value, and it does not affect the fluctuations. The autocorrelation function is given by
\begin{equation}
    \expval{\delta j(0)\delta j(t)} = \expval{\delta j^2} \exp\qty(-\frac{\gamma}{m}t) \, ,
\end{equation}
as shown in the main text.

\section{Edge currents}

The angular momentum of our system can be stored as either spin or orbital angular momenta. Denoting $J, L$ and $S$ as the total, orbital and spin angular momenta of the system, respectively, we express the total angular momentum a sum of orbital and spin angular momenta. 
\begin{equation}
\begin{aligned}
    J&=\sum_{i\alpha} \textbf{r}_{i\alpha}\times \textbf{p}_{i\alpha} \\
    &=\sum_i\frac{\textbf{r}_{i1}+\textbf{r}_{i2}}{2}\times(\textbf{p}_{i1}+\textbf{p}_{i2}) +\frac{\textbf{r}_{i1}-\textbf{r}_{i2}}{2}\times(\textbf{p}_{i1}-\textbf{p}_{i2})\\
    &= L+S.
\end{aligned}
\end{equation}

In the dilute limit, where collisions between dimers are rare, approximately all of the injected angular momentum is stored in and dissipated through the spin degree of freedom, with a spin steady-state average value of $\expval{S}/A =\rho\tau_a /2\gamma$. However, as the system's density increases, collisions between the dimers act to transfer spin angular momentum into orbital angular momentum. The relevant time scales are the relaxation time $\tau_\text{relax}\propto m/\gamma$ and the mean free time $\tau_\text{coll}\propto 1/\rho \sigma\sqrt{T}$, and the transfer between spin and orbital angular momenta takes place when $\tau_\text{coll}\lesssim\tau_\text{relax}$.

\begin{figure}
    \centering
    \includegraphics[width=0.9\columnwidth]{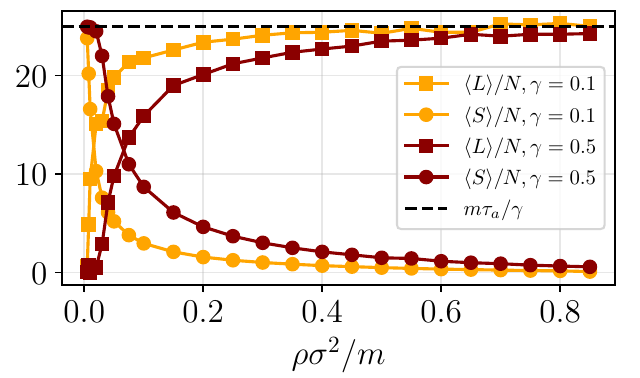}
    \caption{Partitioning of angular momentum in the circular confinement as a function of density. In the dense limit, most of the average angular momentum is stored as  orbital angular momentum due to the collisions between the dimers.}
    \label{fig:angular_momentum_transfer}
\end{figure}

 \subsection{Mean behavior}

\begin{figure}[b]
    \centering
    \includegraphics[width=0.7\columnwidth]{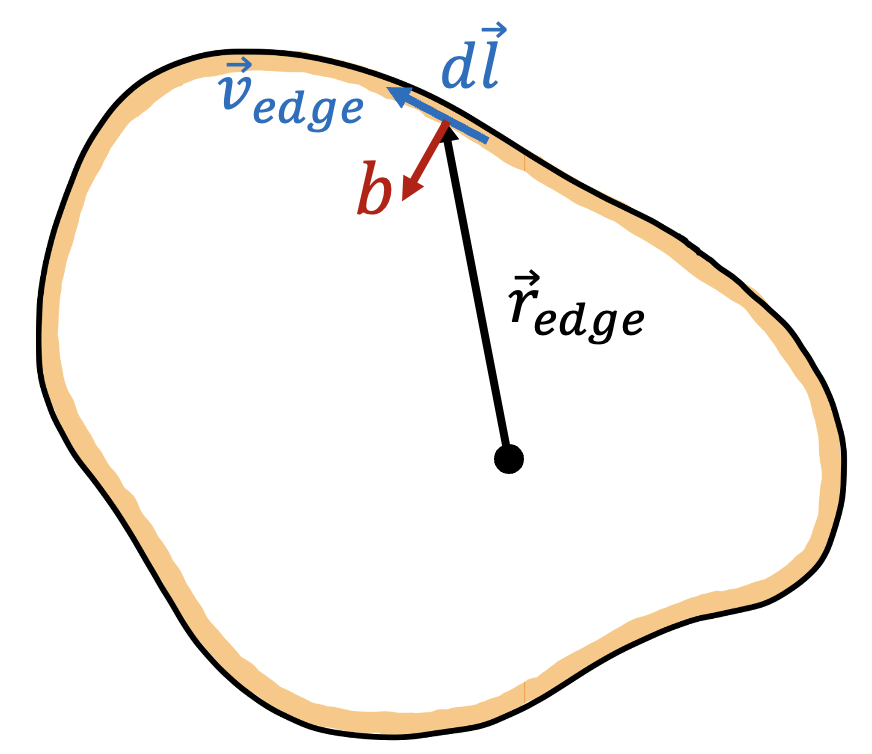}
    \caption{Schematic of the localized velocity field.}
    \label{fig:schematic_edge_velocity}
\end{figure}

It is instructive to start with understanding the mean behavior of the orbital angular momentum. Numerically, we find that $\expval{L}/ \expval{S} \sim \mathcal{O}(100)$ in dense systems, see Fig.~\ref{fig:angular_momentum_transfer}, which is an observation that can be understood intuitively. In the limit of dense systems, the dimers' spin becomes frustrated and unable to build up, transferring approximately all of the injected angular momentum into orbital angular momentum. We then express 
\begin{equation}
    \expval{L}=\int \textbf{r}\times \rho(\textbf{r})\textbf{v}(\textbf{r})  \;dA,
\end{equation}
where $\rho(\textbf{r})$ and $\textbf{v}(\textbf{r})$ are the mean density and velocity fields at position $\textbf{r}$, respectively, and $dA$ is an area element. Since our model is isotropic and homogeneous in the bulk \cite{Hargus2020}, there cannot be a non-zero average flow field in the bulk of the system. However, the isotropy and homogeneity break down close to the boundary of the system. Therefore, the mean velocity field must be concentrated close to the boundary and this localization can be quantified through hydrodynamic length scales, as shown in the next section. The integral therefore can be written as
\begin{equation}
\begin{aligned}
    \expval{L}&\approx\int \textbf{r}_\text{edge}(l)\times \rho(b)v_\text{edge}(b)  \hat{\mathbf{e}}_t \;dl \; db,
\end{aligned}
\end{equation}
where $\textbf{r}_\text{edge}$ is the position vector of a point along the boundary of the system, parameterized by the length along the boundary $l$, $\textbf{v}_\text{edge}=v_\text{edge}\hat{\mathbf{e}}_t$ is the edge velocity pointing along the tangential direction to the wall $\hat{\mathbf{e}}_t$, whose magnitude $v_\text{edge}$ is assumed to depend only on the normal distance $b$ away from the boundary, see Fig.~\ref{fig:schematic_edge_velocity}.  The same assumption applies to the density profile, $\rho=\rho(b)$. These assumptions are  valid in the limit of small curvatures, where the local geometry of the boundary is flat. This approximation allows us to factorize the integral into two factors
\begin{equation}
\begin{aligned}
     \expval{L}&= \qty(\int \rho(b)v_\text{edge}(b)\;db)\qty(\int \textbf{r}_\text{edge}(l) \times  d\textbf{l} )\\
    &= 2\expval{I} A,
\end{aligned}
\end{equation}
where we define the mean current in the continuum limit as
\begin{equation}
    \expval{I}=\qty(\int \rho(b)v_\text{edge}(b)\;db).
\end{equation}
Therefore, in steady state, the mean current in dense systems is directly proportional to the total mean angular momentum density, $\expval{I}=\expval{j}/2= \rho \tau_a /4\gamma$,  as in the main text.

\subsection{Edge current fluctuations}

While the mean behavior of the edge current is understood easily in relation to the mean total angular momentum density, the fluctuations in the edge current cannot be easily related to the fluctuations of the total angular momentum. Despite the localization of the mean edge current to the edge of the sample, the fluctuations are present in the bulk. Therefore, we need to develop the stochastic equation of motion of the edge current independently. This task is difficult in general, but we show it here for two simple geometries, circular confinement and parallel plates confinement.

\textbf{Parallel plates confinement.} We define the instantaneous edge current in the parallel plates confinement as
\begin{equation}
    I(t)=-\sum_{i\alpha} \frac{\textbf{p}_{i\alpha}(t)\cdot \hat{\mathbf{e}}_{x} }{L_x} \frac{y_{i\alpha}}{|y_{i\alpha}|},
\end{equation}
where $\hat{\mathbf{e}}_x$ is the unit vector pointing along the positive $x$-axis, $L_x$ is the length of the plates, and the factor $y_{i\alpha}/|y_{i\alpha}|$ is an indicator function that switches sign as the particle goes from the top half to the lower half and vice versa. The equation of motion for the current can be derived from the underlying Langevin equations
\begin{equation}
    \begin{aligned}
        \frac{d}{dt}I=&-\frac{\gamma}{m} I-\frac{1}{L_x} \sum_{i\alpha}\left[\textbf{F}_{i\alpha}^c+\textbf{F}_{i\alpha}^a \right ]\cdot \hat{\mathbf{e}}_x \frac{y_{i\alpha}}{|y_{i\alpha}|} \\
        &+\sum_{i\alpha}\sqrt{2\gamma \kB T}\boldsymbol{\eta}_{i\alpha}(t)\cdot \hat{\mathbf{e}}_x \frac{y_{i\alpha}}{|y_{i\alpha}|}.
    \end{aligned}
\end{equation}
The sum over noises can be evaluated using the central limit theorem
\begin{equation}
    \begin{aligned}
        -\frac{\sqrt{2\gamma \kB T}}{L_x}\sum_{i\alpha} \pm \eta_{i\alpha}^{x}&=\frac{\sqrt{4\gamma \kB T   N}}{L_x} \zeta(t)\\
        &= \sqrt{2\gamma \kB T \frac{\rho}{m}\frac{L_y}{L_x}} \zeta(t),
    \end{aligned}
\end{equation}
where we absorbed the negative sign into the random variables $\eta_{i\alpha}$ and used $N=\rho L_xL_y/2m$. $\zeta(t)$ is a Gaussian white noise with zero mean and unit variance. 

\begin{figure}
    \centering
    \includegraphics[width=0.75\columnwidth]{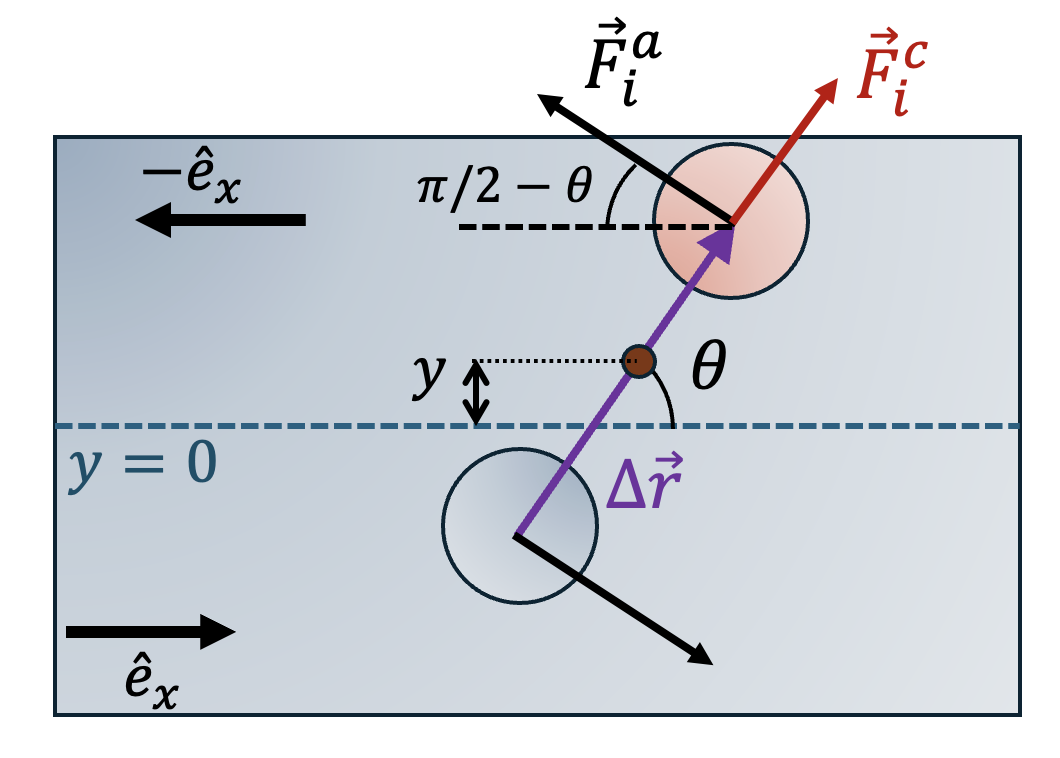}
    \caption{Geometric construction of the contribution from forces close to $y=0$. }
    \label{fig:parallel forces schematic}
\end{figure}

The sum over conservative and active forces can be simplified greatly by realizing that only terms where one monomer is at $y>0$ and the other is at $y<0$ contribute to this sum.  While these contributions are fluctuating in nature, their net contribution is well approximated by their mean for large systems, as the fluctuations from the Langevin thermostat come from all the monomers, dominating over the force fluctuations due to monomers in the vicinity of $y=0$. To calculate the mean contribution, we geometrically construct a mean-field approximation. Let the vector $\Delta \textbf{r}_{pq}$ connecting two interacting monomers make an angle $\theta_{pq}$ with the $x$-axis. Their forces will contribute to the sum as long as $|y_{pq}|\leq \frac{\Delta r_{pq}}{2}\sin(\theta_{pq})$, where $y_{pq}$ is the $y$-coordinate of their center of mass, see Fig.~\ref{fig:parallel forces schematic}.
 In the case of active forces, we may approximate $\Delta r_{pq}=\sigma$, and the average total contribution can be approximated by the isotropy of the system
\begin{equation}
\begin{aligned}
    \frac{2}{L_x}&\sum_{|y_i|\leq \sigma \sin(\theta_i)/2} F^a_i \sin(\theta_i)\\
    &\approx \frac{2}{L_x} \int_0^\pi \frac{d\theta}{\pi} \int_{-\sigma \sin(\theta)/2}^{\sigma \sin(\theta)/2} \frac{\rho L_x}{2m}F^a_i \sin(\theta) dy\\
    &= \frac{\rho}{2m}\tau_a.
\end{aligned}
\end{equation}
The mean contribution from conservative forces can be calculated in a similar fashion, while $\Delta r_{pq}$  follows a wider distribution, its independence from the angular distribution is all that is necessary to see that the mean contribution vanishes
\begin{equation}
\begin{aligned}
    \frac{2}{L_x} &\sum_{|y_{pq}|\leq \Delta r \sin(\theta_{pq})/2} F^c_{pq} \cos(\theta_{pq})\\
    &\propto \int_{0}^\pi \sin(\theta) \cos(\theta) d\theta =0.
\end{aligned}
\end{equation}
This result is expected, since conservative forces do not spontaneously lead to a mean current in steady-state.

Combining all of these results, we reach a stochastic differential equation for the edge current
\begin{equation}
    \frac{d}{dt}I=-\frac{\gamma}{m}I+\frac{\rho}{2m}\tau_a +\sqrt{2\gamma \kB T \frac{\rho}{m}\frac{L_y}{L_x}} \zeta(t).
\end{equation}
The steady-state statistics of this process is mathematically equivalent to those of the total angular momentum density.

\begin{figure}
    \centering
    \includegraphics[width=0.75\columnwidth]{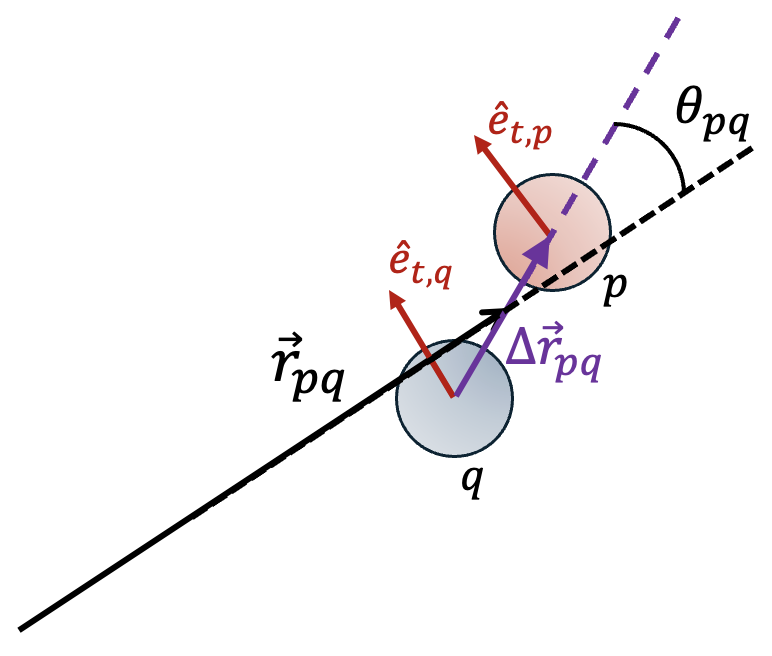}
    \caption{Geometric construction of the contribution from forces in the circular confinement. }
    \label{fig:circular ofrces}
\end{figure}

\textbf{Circular Confinement.} We define the instantaneous edge current in the circular confinement as
\begin{equation}
    I(t)=\sum_{i\alpha} \frac{\textbf{p}_{i\alpha}(t)\cdot \hat{\mathbf{e}}_{t,i\alpha}}{2\pi R},
\end{equation}
where $\hat{\mathbf{e}}_{t,i\alpha}$ is the tangential unit vector of the $i\alpha$ monomer, defined as $\hat{\mathbf{e}}_{t,i\alpha}=\hat{\mathbf{e}}_z\times \textbf{r}_{i\alpha}/r_{i\alpha}$, and $R$ is the radius of the circular confinement.  The equation of motion can be derived in a similar fashion to that used for the parallel-plate confinement
\begin{equation}
    \dot{I}=-\frac{\gamma}{m}I + \frac{1}{2\pi R} \sum_{i\alpha}\left[\textbf{F}_{i\alpha}^c+\textbf{F}_{i\alpha}^a+\sqrt{2\gamma \kB T}\boldsymbol{\eta}_{i\alpha}(t)\right]\cdot \hat{\mathbf{e}}_{t,i\alpha}, 
\end{equation}
where we ignored the time derivative of the tangential unit vector, as its mean vanishes and its fluctuations are negligible compared to the fluctuations due to noise.  The sum over noise forces can be done via the central limit theorem, giving 
\begin{equation}
\begin{aligned}
    \frac{\sqrt{2\gamma \kB T}}{2\pi R} \sum_{i\alpha} \eta_{i\alpha}^t&= \frac{\sqrt{4\gamma \kB T  N}}{2\pi R}    \zeta(t)\\
    &=\sqrt{\frac{1}{2\pi} \gamma \kB T \frac{\rho}{m}} \zeta(t).
\end{aligned}
\end{equation}
In order to calculate the mean contribution from the forces, we consider two interacting monomers, with separation $\Delta \textbf{r}_{pq}=\Delta r_{pq} \hat{\mathbf{e}}_{s,pq}$, where $\hat{\mathbf{e}}_{s,pq}$ is the unit vector pointing from $q$ to $p$. The position vector from the center of the circle to their center of mass is $\textbf{r}_{pq}=r_{pq} \hat{\mathbf{e}}_{r,pq}$, and the angle between $\textbf{r}_{pq}$ and $\Delta \textbf{r}_{pq}$ is $\theta_{pq}$, see Fig.~\ref{fig:circular ofrces}. It is convenient to express 
\begin{equation}
    \frac{\textbf{r}_{p}}{r_p}-\frac{\textbf{r}_q}{r_q}\approx \frac{\Delta r_{pq}}{r} \qty[\hat{\mathbf{e}}_{s,pq}-\hat{\mathbf{e}}_{r,pq} \cos(\theta_{pq})].
\end{equation}
In the case of active forces, we can write the mean total contribution as
\begin{equation}
    \begin{aligned}
       & \frac{1}{2\pi R}\sum_i \textbf{F}^a_{i1} \cdot (\hat{\mathbf{e}}_{t,i1}-\hat{\mathbf{e}}_{t,i2})\\
        &\approx \frac{1}{2\pi R} \sum_{i} F^a_i (\hat{\mathbf{e}}_z \times \hat{\mathbf{e}}_{s,i}) \cdot\qty(\hat{\mathbf{e}}_z \times \frac{\Delta r_i}{r_i}(\hat{\mathbf{e}}_{s,i}-\hat{\mathbf{e}}_{r,i}\cos(\theta_i)))\\
        &=\frac{\tau_a}{2\pi R} \sum_i \frac{\sin^2(\theta_i)}{r_i}\\
        &\approx \frac{\tau_a}{2\pi R}\frac{N}{\pi R^2} \int_{0}^{2\pi} \sin^2(\theta) d\theta \int_0^R \frac{1}{r}r dr= \frac{\rho}{4m} \tau_a.
    \end{aligned}
\end{equation}

\begin{figure}[t]
    \centering
    \includegraphics[width=\columnwidth]{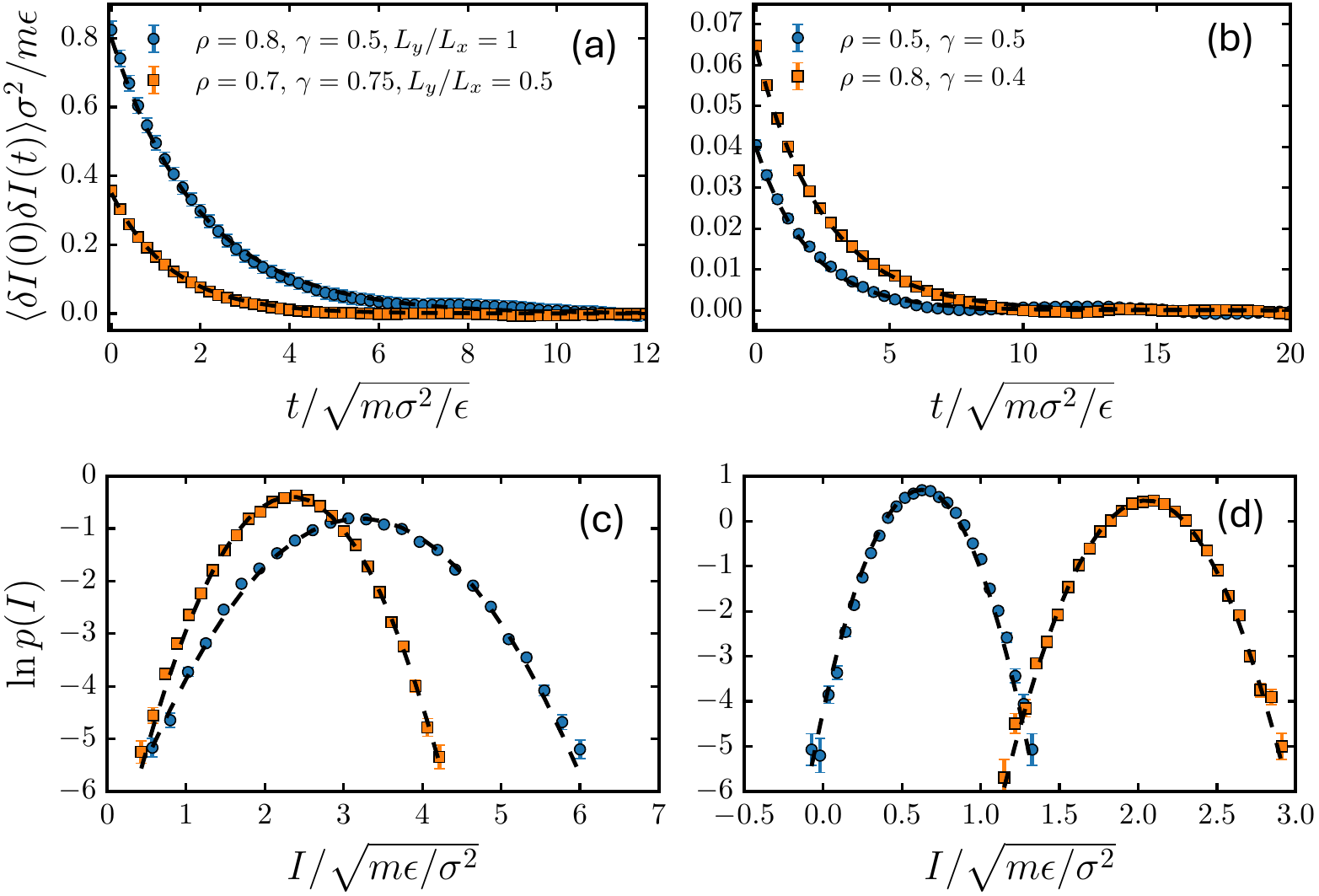}
    \caption{Autocorrelation function of the edge current for the rectangular (a) and circular (b) geometries at different system parameters. Probability distribution of the edge current for the rectangular (c) and circular (d) geometries corresponding to the parameters in (a) and (b), respectively. The dashed black lines are the theoretical predictions. }
    \label{fig:current statistics}
\end{figure}

In the case of conservative forces, a similar calculation can be done, where we find the mean contribution to vanish
\begin{equation}
\begin{aligned}
     \frac{1}{2\pi R} &\sum_{pq} \textbf{F}^c_{pq}\cdot (\hat{\mathbf{e}}_{t,p}-\hat{\mathbf{e}}_{t,q}) \propto \sum_{pq} \frac{\sin(\theta_{pq})\cos(\theta_{pq})}{r_{pq}}\\
     &\propto \int_0 ^{2\pi} \sin(\theta)\cos(\theta)d\theta =0,
\end{aligned}
\end{equation}
which is also an expected result. The fluctuating contributions from the forces are present everywhere in the system. However, as we increase the system size, these contributions vanish as $\hat{\mathbf{e}}_{t,p}-\hat{\mathbf{e}}_{t,q}\to 0$ for large $r_{pq}$, while the fluctuations supplied by the noise forces maintain their strength, which justifies ignoring the fluctuating contributions from the forces.
Combining all of these results, we find
\begin{equation}
    \frac{d}{dt}I=-\frac{\gamma}{m}I+\frac{\rho}{4m}\tau_a +\sqrt{\frac{1}{2\pi}\gamma \kB T \frac{\rho}{m}}\zeta(t),
\end{equation}
as in the main text. Figure \ref{fig:current statistics}
 shows the numerical results for the autocorrelation function and steady-state distribution for the edge current, demonstrating good agreement with the predictions. The rectangular geometries were of dimensions $(L_x,L_y)=(400\sigma,400\sigma)$ and $(300\sigma,600\sigma)$, while the circular confinements were of radius $R=200\sigma$ for $\rho=0.8 m\sigma^{-2}$ and $R=250\sigma$ for $\rho=0.5 m\sigma^{-2}$.

\section{Application to Other Chiral Systems}

In this section, we apply the global angular momentum balance formalism to derive expressions for the mean edge current for an extension of our model as well as other chiral systems of importance. The derivations will assume that the liquids are two dimensional and homogeneous, that the edge current is sufficiently localized compared to the system size, and that the boundary's curvature is small compared to the particle size, so that the expression $\expval{L}=2\expval{I}A$ is valid. We assume that all wall forces are short-ranged. 

In all the following models, $m$ is the mass of a single particle, $I$ is the moment of inertia of a single particle with respect to its center of mass, $\gamma$ and $\gamma_r$ are the translational and rotational substrate drag coefficients, $T$ is the bath temperature, $\eta(t)$ and $\zeta(t)$ are Gaussian white noise with zero mean and unit variance, $\textbf{r}_i$ and $\textbf{p}_i$ are the position and  linear momentum vectors of the $i$th particle, and $\theta_i$ and $s_i$ are the director's angle and spin of the $i$th particle. Pairwise central forces are denoted as $\textbf{F}^c$, and wall forces are denoted as $\textbf{F}^\text{w}$. $\rho$ is the mass density, $\expval{I}$ (different from $I$) is the mean edge current measured in units of mass per unit time, and $A$ is the area.

\subsection{Active Dumbbells with Distributed Torques, Frictional Walls and Residual Spin}

We extend our main model described by equation \ref{Dumbbell Model} by allowing the active torque to depend on the dimer, such that $F_{i\alpha}^a(t)d_i(t)=\tau_a^i$. In particular, the sign of $\tau_a^i$ can change. Moreover, the wall can be endowed with a tangential force, such as friction. Then for the total angular momentum $J$ defined in equation \ref{total angular momentum definition}, we find
\begin{equation}
    \dot{\expval{J}}=-\frac{\gamma}{m} \expval{J} +N\expval{\tau_a}+\sum_{i\alpha} \expval{\textbf{r}_{i\alpha}\times \textbf{F}_{i\alpha}^\text{w}},
\end{equation}
where the sum over all the active torques is replaced by their mean. Following earlier derivations, the wall term can be decomposed into contributions from the perpendicular (central) and tangential forces
\begin{equation} \label{Frictional Wall}
\begin{aligned}
    \sum_{i\alpha} \expval{\textbf{r}_{i\alpha}\times \textbf{F}_{i\alpha}^\text{w}}&\approx \int \textbf{r}\times d\textbf{F}^\text{w}\\
    &= \int \textbf{r}\times \qty[B(\textbf{r}) d\textbf{l}\times \hat{z}+ \Gamma(\textbf{r}) d\textbf{l}]\\
    &\approx B_o \int \textbf{r}\times (d\textbf{l}\times \hat{z}) + \Gamma_o \int \textbf{r}\times d\textbf{l}\\
    &= 2\Gamma_o A,
\end{aligned}
\end{equation}
where in the second line we introduced a function $\Gamma(\textbf{r})$ to quantify the strength of the tangential wall forces, which in principle depends on the density and other parameters according to the wall model used. The crucial step is approximating it as a constant in the third line with the same justification used for $B(\textbf{r})$. This shows that the contribution from the tangential wall forces scales with the area. Then, decomposing $\expval{J}=\expval{L}+\expval{S}$, we can write an expression for the mean edge current in steady-state
\begin{equation}
    \expval{I}=\frac{\rho \expval{\tau_a}}{4\gamma}-\frac{\expval{S}}{2A}+\frac{m}{\gamma}\Gamma_o.
\end{equation}
We note that the constant active torque in our original model is simply replaced by its mean, and the residual spin density can be neglected for the densities we consider. The tangential wall forces can either empower or hinder the edge current depending on the sign of $\Gamma_o$. For a reasonable friction model, we expect $\Gamma_o$ to be odd in $\expval{\tau_a}$ and have the opposite sign. In particular, if $\expval{\tau_a}=0$ for a mixed system, then $\expval{I}=0$.

\subsection{Dipoles Driven by a Planar Rotating External Field}

One practical application of our method is to electric or magnetic dipoles driven by a rotating external field \cite{soni2019odd}. Consider $N$ dipoles sedimented on a two dimensional plane. For simplicity, their dipole moments $\boldsymbol{\mu}_i$ are assumed to have the same magnitude $\mu$ and lie on the two dimensional plane. Each interacts with the typical three-dimensional dipole-dipole interaction. In addition to dipole-dipole interactions, the dipoles interact with central forces provided by a pairwise interaction potential. The dipoles are subject to a rotating, uniform external magnetic field $\textbf{B}(t)={B}  \, \qty(\cos(\omega_\text{B}t),\sin(\omega_\text{B}t))$ with amplitude ${B}$ and frequency $\omega_\mathrm{B}$. We assume the dipoles evolve according to under-damped Langevin equations
\begin{equation}
    \begin{aligned}
        \dot{\textbf{r}}_i&=\textbf{p}_i/m,\\
        \dot{\theta}_i &=s_i/{I},\\
        \dot{\textbf{p}}_i&= -\frac{\gamma}{m}\textbf{p}_i+\textbf{F}_{i}^c+\textbf{F}_{i}^\text{dd}+\sqrt{2\gamma \kB T} \boldsymbol{\eta}_i(t),\\
        \dot{s}_i&= -\frac{\gamma_r}{{I}} s_i +\tau_i^\text{dd}+ \boldsymbol{\mu}_i\times \textbf{B}(t)+\sqrt{2\gamma_r\kB T}\zeta_i(t),
    \end{aligned}
\end{equation}
 where $\textbf{F}^\text{dd}_i$ and $\tau^\text{dd}_i$ are the net dipole-dipole force and torque acting on the $i$th dipole. The total angular momentum of this system will be the sum of orbital and spin angular momenta
\begin{equation}
    J=L+S=\sum_i \textbf{r}_i\times \textbf{p}_i + s_i.
\end{equation}
The evolution of the total angular momentum can be found from the underlying equations of motion, leading to
\begin{equation}
\begin{aligned}
    \expval{\dot{J}} = -\frac{\gamma}{m}&\expval{L}-\frac{\gamma_r}{{I}} \expval{S}+\expval{\sum_{i}\textbf{r}_i\times \textbf{F}_i^\text{dd}+\tau_i^\text{dd}} \\&+ \expval{\sum_{i}\boldsymbol{\mu}_i\times \textbf{B}(t)}.
\end{aligned}
\end{equation}
The contribution from the dipole-dipole forces and torques cancels since they conserve the total angular momentum. Therefore, in steady state, we have the relationship
\begin{equation}
    \frac{\gamma}{m}\expval{L}=\expval{\sum_i\boldsymbol{\mu}_i\times \textbf{B}(t)}-\frac{\gamma_r}{{I}}\expval{S},
\end{equation}
 which can be used to find the mean edge current as a function of the average spin and injected torque through the external field. 
 
 A particularly simple limit to consider is the phase-locked regime, where the angular frequency of the dipoles matches that of the external field \cite{soni2019odd}. In this limit, the phase lag between the dipoles and the external magnetic field is fixed, $\phi_i=\omega_\text{B} t-\theta_i$, resulting in
\begin{equation}
    \frac{\gamma}{m}\expval{L}=N\mu B\expval{ \sin(\phi)}  -\gamma_r N\omega_\text{B},
\end{equation}
leading to an expression of the mean edge current
\begin{equation}
    \expval{I} = \frac{\rho}{2\gamma} \qty[\mu B\expval{ \sin(\phi)} -\gamma_r\omega_\text{B}].
\end{equation}

Note that in the dilute limit, where dipole-dipole interactions are neglected, $\mu B \expval{\sin(\phi)} =\gamma_r\omega_\text{B}$, resulting in a zero edge current, as expected. To gain further intuition behind the result, consider the spin equation of motion for a tagged particle in the phase-locked, steady state limit
\begin{equation}
    -\gamma_r \omega_\text{B} +\mu B \expval{\sin(\phi)}+\expval{\tau^\text{dd}}=0,
\end{equation}
where $\expval{\tau^\text{dd}}$ is the mean torque exerted by the surrounding dipoles on the tagged dipole. For an equilibrium distribution, $\tau^\text{dd}=0$. In a non-equilibrium distribution, induced by $\omega_\text{B}\neq0$, the symmetry of the system requires $\expval{\tau^\text{dd}}$ be odd in $\omega_\text{B}$. Moreover, since the dipole-dipole interactions are short ranged, we expect their mean on a tagged particle to scale linearly with density, leading to an expression for the mean edge current of the form
\begin{equation}
    \expval{I} = -\frac{\rho}{2\gamma} \expval{\tau^\text{dd}}\equiv A_2\frac{\rho^2}{2\gamma}  \omega_\text{B} + \mathcal{O}(\omega_\text{B}^3),
\end{equation}
where $A_2$ is a parameter that is independent of $\omega_\text{B},{B}$ and $
\rho$, but can depend on other system parameters, as described below. Therefore, in the small driving frequency regime, where the phase-locking is expected to hold, the edge current is expected to rise linearly with the driving frequency, which agrees with the observations in Ref. \cite{soni2019odd}. 

To link this dependence more concretely to the system properties, we define the pair distribution function  $g(r,\alpha)$, where $r$ is the separation distance between the pair, and $\alpha$ is the angle between the separation vector and a reference axis, taken to be the common dipole moment direction. The $g(r,\alpha)$ is known to have non-trivial anisotropic components in chiral active systems \cite{Poggioli2023_Mobility}, so we expand it in harmonics
\begin{equation}\label{harmonic expansion}
    g(r,\alpha)=g_0(r)+\sum_{n=1} \qty[a_n(r) \sin(n\alpha)+b_n (r)\cos(n \alpha)].
\end{equation}
In the bulk, the symmetry $g(r,\alpha)=g(r,\alpha+\pi)$ sets the odd harmonic terms to zero. Moreover, the implicit dependence on $\omega_\text{B}$ through $g(r,\alpha;\omega_\text{B})=g(r,-\alpha;-\omega_\text{B})$ guarantees that $g_0$ and the $b_n$'s are even in $\omega_\text{B}$, while the $a_n$'s are odd in it. We can express the mean dipole-dipole torque in terms of an integral over $g(r,\alpha)$
\begin{equation}
    \expval{\tau^\text{dd}}= \frac{\rho}{m} \int_0^\infty rdr \int_0^{2\pi} d\alpha \; g(r,\alpha) \tau^\mathrm{dd}(r,\alpha),
\end{equation}
\noindent where $\rho/m$ is the number density, and $\tau^\text{dd}(r,\alpha)$ is the torque between two aligned dipoles, and is given by 
\begin{equation}
    \tau^\text{dd}(r,\alpha)= \frac{3\mu^2}{2r^3}\sin(2\alpha).
\end{equation}
Performing the angular integral, only the $a_2$ harmonic survives, giving 
\begin{equation}
    \expval{\tau^\text{dd}}= \frac{3\pi \mu^2\rho}{2m} \int_0^\infty \frac{a_2(r)}{r^2}dr\equiv -A_2\rho \omega_\text{B}+\mathcal{O}(\omega_\text{B}^3) 
\end{equation}
which from symmetry should scale as $\omega_\mathrm{B}$.

\subsection{Active Frictional Disks}

Consider $N$ identical rigid disks of radius $R$. The disks interact with each other via central and frictional contact forces, and they are each powered by an active torque $\tau_a$ measured around their individual center of mass. The disks are confined by a frictional wall. The evolution of the system is described by under-damped Langevin equations
\begin{equation}
    \begin{aligned}
        \dot{\textbf{r}}_i&=\textbf{p}_i/m,\\
        \dot{\theta}_i&=s_i/{I},\\
        \dot{\textbf{p}}_i&=-\frac{\gamma}{m}\textbf{p}_i +\textbf{F}_i^c+\textbf{F}_i^f+\textbf{F}_i^\text{w}+\sqrt{2\gamma \kB T}\boldsymbol{\eta}_i(t),\\
        \dot{s}_i&=-\frac{\gamma_r}{{I}}s_i +\tau_i^f+\tau_a+\tau_i^\text{w}+\sqrt{2\gamma_r \kB T}\zeta_i(t),
    \end{aligned}
\end{equation}
where $\textbf{F}_i^f$ and $\tau_i^f$ are the total frictional force and torque acting on the $i$'th disk due to contact with other disks, while $\tau_i^\mathrm{w}$ is the frictional torque exerted by the wall.
The total angular momentum of the system is given by $J=\sum_i \textbf{r}_i\times\textbf{p}_i+s_i$, and the evolution of its mean is given by
\begin{equation}
\begin{aligned}
    \dot{\expval{J}}=&- \frac{\gamma}{m} \sum_i \textbf{r}_i \times \textbf{p}_i -\frac{\gamma_r}{\mathrm{I}}\sum_i s_i+\sum_i \textbf{r}_i\times \textbf{F}_i^f + \tau_i^f \\
    &+N\tau_a +\sum_i \textbf{r}_i\times \textbf{F}^\text{w}_i +\tau_i^\text{w}.
\end{aligned}
\end{equation}
The contribution from the pairwise contact frictional forces and torques vanishes as they conserve the total angular momentum. Moreover, the sum over the wall forces can be computed similarly to equation \ref{Frictional Wall}. However, the sum over the wall torques can be neglected for a large system since it only scales with the perimeter and not the area. Rearranging the equation at steady-state,
\begin{equation}
    \frac{\gamma}{m}\expval{L}= N\tau_a -\frac{\gamma_r}{\mathrm{I}}\sum_is_i +2\Gamma_o A,
\end{equation}
where $\Gamma_o<0$ is the characteristic friction strength. Assuming that disks in the bulk spin at a uniform angular speed $\Omega$, we can further simplify the equation and find an expression for the edge current
\begin{equation}
    \expval{I}=\frac{\rho}{2\gamma}\qty(\tau_a - \gamma_r\Omega ) +\frac{m}{\gamma}\Gamma_o.
\end{equation}

\subsection{Chiral Active Brownian Particles}

Another example of interest is chiral active Brownian particles. Consider $N$ particles of mass $m$ propelled by an active force $\textbf{F}_i^a=F_a \hat{u}_i(\theta_i)$, where $\hat{u}_i=\cos(\theta_i)\hat{e}_x+\sin(\theta_i)\hat{e}_y$ is the director of the active force. The central interactions between the particles are derived from a potential $U(r_{12})$. We assume the system to be confined by a frictionless wall. The calculation can be extended to the case of frictional walls as was done in Eq.~\ref{Frictional Wall}. The director evolves according to a stochastic equation of motion
\begin{equation}
    \dot{\theta}_i=\Omega+\sqrt{2D_r}\zeta_i(t),
\end{equation}
where $\Omega$ is the deterministic angular frequency, $D_r$ is the rotational diffusion. The translational degrees of motion evolve according to the underdamped Langevin equations
\begin{equation}
\begin{aligned}
    \dot{\textbf{r}}_i &=\textbf{p}_i/m,\\
    \dot{\textbf{p}}_i&=-\frac{\gamma}{m}\textbf{p}_i +\textbf{F}_i^c+\textbf{F}_i^\text{w}+F_a \hat{u}(\theta_i) +\sqrt{2\gamma \kB T}\boldsymbol{\eta}_i(t),\\
\end{aligned}
\end{equation}
Since this model does not have an explicit spin degree of freedom, all injected angular momentum is stored and dissipated as orbital angular momentum $L=\sum_i \textbf{r}_i\times \textbf{p}_i$, whose average evolution is given by
\begin{equation}
    \expval{\dot{L}}=-\frac{\gamma}{m}\expval{L}+\expval{\sum_i \textbf{r}_i \times \textbf{F}_i^a} +\expval{\sum_i \textbf{r}_i \times \textbf{F}_i^\text{w}},
\end{equation}
where the contribution from pairwise central forces vanishes due to Newton's third law. The contribution from the conservative wall forces can be approximated as zero as shown previously. In steady state
\begin{equation}
\begin{aligned}
    \frac{\gamma}{m}\expval{L}&= \expval{\sum_i \textbf{r}_i \times \textbf{F}_i^a}\\
    &\equiv N \expval{\tau_a},\\
\end{aligned}
\end{equation}
and the mean edge current is then given by
\begin{equation}
    \expval{I}= \frac{\rho}{2\gamma} \expval{\tau_a}
\end{equation}
just as in the main text, however with an averaged torque replacing the deterministic one.

To gain further intuition behind this result, we calculate the mean active torque $\expval{\tau_a}$ in the dilute limit, where we can neglect the pairwise interactions, leading to
\begin{equation}
    \begin{aligned}
        \dot{\textbf{p}}_i&=-\frac{\gamma}{m}\textbf{p}_i +F_a \hat{u}(\theta_i) +\sqrt{2\gamma \kB T}\boldsymbol{\eta}_i(t),\\
        \dot{\theta}_i&=\Omega+\sqrt{2D_r}\zeta_i(t).
    \end{aligned}
\end{equation}
The active torque is $\tau_i^a(t)=F_a \qty[x_i \sin(\theta_i)-y_i \cos(\theta_i)]$, which with the help of the complex coordinates $\mathtt{z}_i=x_i+\mathrm{i}y_i$ and $w_i=e^{\mathrm{i}\theta_i}$ can be expressed as $\tau_i^a= F_a \text{Im}(\bar{\mathtt{z}}_i w_i)$. The equation of motion for $\mathtt{z}_i$ is
\begin{equation}
    m \ddot{\mathtt{z}}_i= -\gamma \dot{\mathtt{z}}_i +F_a w_i(\theta_i) +\sqrt{2\gamma \kB T} (\eta_{i,x}+\mathrm{i}\eta_{i,y}),
\end{equation}
whose formal particular solution is given by
\begin{equation}
\begin{aligned}
    \mathtt{z}_i(t)=F_a&\int_0^t G(t-t') w_i(\theta_i(t'))dt'\\
    &+\sqrt{2\gamma \kB T}\int_0^t (\eta_{i,x}+\mathrm{i}\eta_{i,y}) dt',
\end{aligned}
\end{equation}
where $G(t)$ is the response function given by
\begin{equation}
    G(t)=\frac{1}{\gamma}\qty[1-e^{-\gamma t/m}].
\end{equation}
From this formal expression, we can calculate the needed expectation value
\begin{equation}
    \begin{aligned}
        \expval{\bar{\mathtt{z}}(t) w(t)}&= F_a \int_0^t G(t-t') \expval{\bar{w}(t')w(t)} \;dt'\\
        &=F_a \int_0^t G(t-t') \expval{e^{\mathrm{i}(\theta(t)-\theta(t'))}}dt'\\
        &=F_a \int_0^t G(t-t') \exp\qty[(-D_r+\mathrm{i}\Omega )(t-t')]dt'\\
        &= F_a \int_0^t G(\tau) \exp\qty[(-D_r+\mathrm{i}\Omega )\tau],\\
    \end{aligned}
\end{equation}
where in going from the second to third lines, we used that $\theta(t)$ has Gaussian statistics with a constant drift. The steady-state average active torque can then be found as
\begin{equation}
\begin{aligned}
        \expval{\tau_a}&=\lim_{t\to \infty} F_a \text{Im}(\expval{\bar{\mathtt{z}}_i(t) w_i(t)})\\
        &=F_a^2\frac{\Omega(2mD_r+\gamma)}{[m(D_r^2-\Omega^2)+\gamma D_r]^2+\Omega^2[2mD_r+\gamma]^2}\\
        &\equiv F_a R_\text{eff} ,
\end{aligned}
\end{equation}
where $R_\text{eff}$ is the effective arm length. 

The averaged torque is odd in $\Omega$ as expected, and scales quadratically with $F_a$, since the arm length scales linearly with $F_a$. One limit of interest is the overdamped limit, where we find
\begin{equation}
    \lim_{m/\gamma\to0}\expval{I} = \frac{\rho}{2\gamma}F_a \lim_{m/\gamma\to0} R_\text{eff}= \frac{\rho}{2} \frac{F_a^2}{\gamma^2}\frac{\Omega}{D_r^2+\Omega^2}.
\end{equation}
This result agrees with the result in Ref. \cite{metzger2026equation} for $F_a=\gamma v_o$ and $\Omega=\omega_o$. Another interesting case is where the distribution of angular frequencies is such that a fraction $p$ rotates with $\Omega$ and a fraction $1-p$ rotates with $-\Omega$, then the edge current is  given by
\begin{equation}
    \expval{I}=(2p-1)\frac{\rho F_a^2}{2\gamma^2} \frac{\Omega}{\Omega^2+D_r^2},
\end{equation}
in the overdamped limit. We expect this setup to qualitatively explain the edge currents in \cite{li2024robust}. In the dense limit, it would be necessary to either numerically calculate the average torque or find an approximation for the effective drag coefficient $\gamma \to \Gamma(\rho)$, which will depend on density. If such an approximation is made, then the final result for the edge current in the overdamped limit should be replaced by
\begin{equation}
    \expval{I}= \frac{\rho}{2} \frac{F_a^2}{\gamma \Gamma(\rho)}\frac{\Omega}{D_r^2+\Omega^2},
\end{equation}
where only one of the $\gamma$'s is replaced, since the $\gamma$ arising from the sum over angular momenta is unaffected by the presence of pairwise central interactions. One reasonable approximation is a linear dependence on density $F_a/\Gamma(\rho)\approx F_a (1-\rho/\rho^*)/\gamma$, which is used for achiral active Brownian particles \cite{Fily2012Athermal,cates2014motility}, where $\rho^*$ is a characteristic density expected to be close to the close-packing density. Following such an approximation, the edge current is expected to rise linearly in the dilute limit, saturating at a finite value of density $\rho^\star/2$, then falling to zero. Since this system breaks parity symmetry, we expect there to be an effective odd mobility \cite{Poggioli2023_Mobility}, and that the replacement of $\gamma$ by $\Gamma(\rho)$ does not account for it. However, we expect this effect to be marginal.

\subsection{Chiral Particles Interacting with Tangential Pairwise Forces}

We finally consider $N$ chiral passive Brownian particles, whose pairwise interactions contain central and tangential forces. The force between two particles $i$ and $j$ takes the form $\textbf{F}_{ij}=\textbf{F}_{ij}^\parallel+\textbf{F}_{ij}^\perp=-\nabla_iU^{\parallel}(r_{ij}) -\hat{z}\times \nabla_i U^\perp(r_{ij})$. Their dynamics evolve according to the under-damped Langevin equations
\begin{equation}
    \begin{aligned}
        \dot{\textbf{r}}_i&=\textbf{p}_i/m,\\
        \dot{\textbf{p}}_i&=-\frac{\gamma}{m}\textbf{p}_i+\sum_{j}\textbf{F}_{ij}+\sqrt{2\gamma\kB T}\boldsymbol{\eta}_i(t).
    \end{aligned}
\end{equation}
Since there are no explicit spin degrees of freedom in this model, angular momentum is stored as orbital angular momentum $L=\sum_i \textbf{r}_i\times\textbf{p}_i$, and its mean evolution is given by
\begin{equation}
    \dot{\expval{L}}= -\frac{\gamma}{m}\expval{L}+\expval{\sum_{ij}\textbf{r}_i\times\textbf{F}_{ij}^\perp},
\end{equation}
where the contribution from central forces vanishes due to Newton's third law. If we assume that the tangential forces abide by Newton's third law, we may rewrite the sum as
\begin{equation}
\begin{aligned}   \expval{\sum_{ij}\textbf{r}_i\times\textbf{F}_{ij}^\perp}&= \frac{1}{2}\expval{\sum_{ij}\textbf{r}_{ij}\times\textbf{F}_{ij}^\perp}\\
    &= \frac{1}{2}\expval{\sum_{ij}\qty(x_{ij}F_{ij}^{y,\perp}-y_{ij}F_{ij}^{x,\perp})}\\
    &\equiv 2A\sigma_\text{odd},
\end{aligned}
\end{equation}
where the difference between the off-diagonal terms of the stress tensor is related to the odd stress. In steady state, the average edge current is then given by
\begin{equation}
    \begin{aligned}
        \expval{I}= \frac{m}{\gamma } \sigma_\text{odd},
    \end{aligned}
\end{equation}
similar to the result obtained in Ref.~\cite{metzger2026equation}. 

Expressing the tangential force as a rotated gradient allows us to rewrite the sum over the torques in a convenient form
\begin{equation}
\begin{aligned}
    \frac{1}{2}\expval{\sum_{ij}\textbf{r}_{ij}\times\textbf{F}_{ij}^\perp}&= -\frac{1}{2} \expval{\sum_{ij} \textbf{r}_{ij} \times \qty(\hat{z}\times \nabla U^\perp (r_{ij}))}\\
    &= -\frac{1}{2} \expval{\sum_{ij} \textbf{r}_{ij}\cdot \nabla U^\perp (r_{ij})}\\
    &\approx -\frac{N^2}{2A} \int_0^\infty  \int_0^{2\pi} r\frac{dU^\perp}{dr} g(r,\alpha)\; r\;d\alpha dr,
\end{aligned}
\end{equation}
where $g(r,\alpha)$ is the pair distribution function. Expanding it in harmonics as in equation \ref{harmonic expansion}, we find that only the $g_0$ term survives since the tangential forces depend only on the distance. Therefore, an equivalent way to express the mean edge current is through
\begin{equation}
    \expval{I}=- \frac{\pi}{2}\frac{\rho^2}{m\gamma} \int_0^\infty r^2 g_0(r) \frac{dU^\perp}{dr} \;dr
\end{equation}
which includes an explicit dependence on the magnitude of the tangential forces as well as an implicit one through potential changes to the $g_0(r)$. 

\subsection{Connection to Hydrodynamics}
For a generic 2D confined fluid whose momentum field evolves according to the equation
\begin{equation}
    \partial_t \rho v_\alpha+\partial_\beta (\rho v_\alpha v_\beta )=\partial_\beta \sigma_{\alpha\beta} -\frac{\gamma}{m} \rho v_\alpha,
\end{equation}
we can write down the equation for orbital angular momentum density $l=\epsilon_{\alpha\beta} r_\alpha \rho v_\beta$
\begin{equation}
    \partial_t l +\partial_\beta (v_\beta l)=-\frac{\gamma}{m} l-\epsilon_{\alpha\beta} \sigma_{\beta \alpha}+\epsilon_{\alpha\beta} \partial_\gamma (r_\alpha \sigma_{\beta\gamma}),
\end{equation}
where the second term on the RHS is twice the odd stress. Integrating over the entire confined region, we find
\begin{equation}
\begin{aligned}
    \frac{d}{dt}L+ \int_A \partial_\beta v_\beta l \; dA=& -\frac{\gamma}{m} L+2\int_A \sigma_\text{odd} \;dA  \\
    &+\int_A \epsilon_{\alpha \beta} \partial_\gamma (r_\alpha \sigma_{\beta\gamma}) \;dA,
\end{aligned}
\end{equation}
where $L=\int_A l \;dA$. Using the divergence theorem
\begin{equation}
\begin{aligned}
    \frac{d}{dt}L + \oint_\mathcal{P} l  \textbf{v} \cdot \hat{n} \; ds =&-\frac{\gamma}{m} L+2\int \sigma_\text{odd} \;dA 
    \\&+\oint_\mathcal{P} \epsilon_{\alpha\beta} r_\alpha \sigma_{\beta \gamma} n_\gamma \;ds.
\end{aligned}
\end{equation}
The second term on the LHS is zero as we assume no particles escape the confinement, $\textbf{v}\cdot \hat{n}=0$.  The last term on the right hand side is the contribution from the wall, which can be treated similarly to Eq.~ \ref{Frictional Wall}. In the case of a frictionless wall, this contribution vanishes. Therefore, in steady state, 
\begin{equation}
    \frac{\gamma}{m} L=2\int_A \sigma_\text{odd} \;dA \approx 2\sigma_\text{odd}^\text{bulk} A,
\end{equation}
\noindent where we approximated the total integral over the odd stress to be its value in the bulk times the total area. This then implies that the orbital angular momentum is given by
\begin{equation}
    L=\frac{2m}{\gamma}\sigma_\text{odd}^\text{bulk}A,
\end{equation}
which leads to an edge current of the form
\begin{equation}
    I=\frac{m}{\gamma}\sigma_\text{odd}^\text{bulk}
\end{equation}
which establishes that the edge current is generically given by the odd stress.

\bibliography{references}